\documentclass[onecolumn,10pt, final]{article}
\usepackage{authblk}
\usepackage{array}
\usepackage{epsfig} 
\usepackage{amsmath}
\usepackage{epstopdf}
\usepackage{setspace}
\usepackage{comment}
\usepackage{stmaryrd}
\usepackage{bm}
\usepackage{subcaption}

\begin{document}

\begin{titlepage}
  \clearpage\thispagestyle{empty}  
  \noindent
  \hrulefill
  \begin{figure}[h!]
   \centering
   \includegraphics[width=2in]{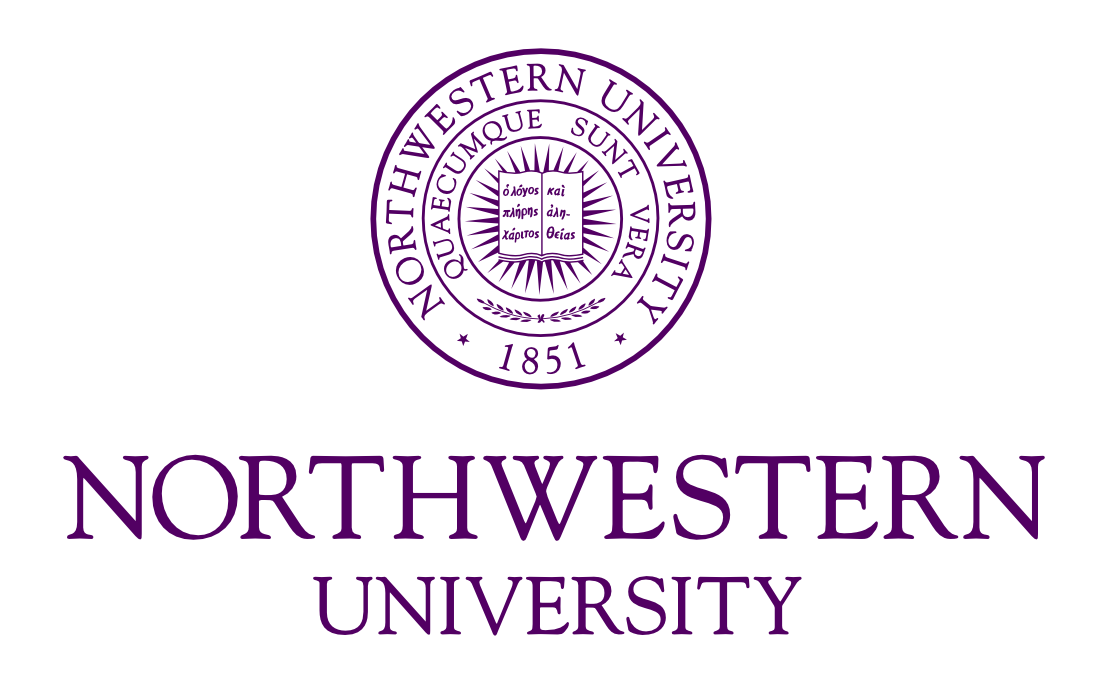}   
  \end{figure}
  \begin{center}
   {
    {
     {\bf Center for Sustainable Engineering of Geological and Infrastructure Materials} \\ [0.1in]
     Department of Civil and Environmental Engineering \\ [0.1in]
     McCormick School of Engineering and Applied Science \\ [0.1in]
     Evanston, Illinois 60208, USA
    }
   }
  \end{center} 
  \hrulefill \\ \vskip 2mm
  \vskip 0.5in
  \begin{center}
   {\large {\bf Isogeometric implementation of high order microplane model for the simulation of high order elasticity, softening, and localization
   }}\\[0.5in]
   {\large {\sc Erol Lale, Xinwei Zhou, Gianluca Cusatis}}\\[0.75in]
   {\sf \bf SEGIM INTERNAL REPORT No. 16/8-587I}\\[0.75in]
  \end{center}
  \vskip 5mm
 \noindent {\footnotesize {{\em Submitted to Journal of Applied Mechanics  \hfill August 2016} }}
 
 \newpage
 \clearpage \pagestyle{plain} \setcounter{page}{1}
 
 \end{titlepage}

\title{Isogeometric implementation of high order microplane model for the simulation of high order elasticity, softening, and localization}

\author[1]{Erol Lale}
\author[2]{Xinwei Zhou}
\author[3]{Gianluca Cusatis\thanks{Corresponding author}}

\affil[1]{\small	Department of Civil Engineering\\
	Istanbul Technical University\\	
	34469 Maslak Istanbul, Turkey}
    
    \affil[2]{\small	ES3, 550 West C St. \\	
		San Diego, CA 92101, USA }
			
\affil[3]{\small Department of Civil and Environmental Engineering,\\	Northwestern University, Evanston, IL 60208, USA.}	
\date{}
\maketitle	

\let\thefootnote\relax\footnote{\footnotesize \textit{Email addresses:} \textbf{lale@itu.edu.tr} (Erol Lale), \textbf{xinwei.zhou@es3inc.com} (Xinwei Zhou), \textbf{g-cusatis@northwestern.edu} (Gianluca Cusatis)}

\doublespacing
\maketitle 
\begin{abstract}
{\it In this paper, a recently developed Higher Order Microplane (HOM) model for softening and localization, is implemented within a isogeometric finite element framework. The HOM model was derived directly from a three dimensional discrete particle model and it was shown to be associated with a high order continuum characterized by independent rotation and displacement fields. Furthermore, the HOM model possesses two characteristic lengths: the first associated with the spacing of flaws in the material internal structure and related to the gradient character of the continuum; and the second associated with the size of these flaws and related to the micro-polar character of the continuum. The displacement-based finite element implementation of this type of continua requires $C^1$ continuity both within the elements and at the element boundaries. This motivated the implementation of the concept of isogeometric analysis which ensures a higher degree of smoothness and continuity. NURBS based isogeometric elements were implemented in a 3D setting, with both displacement and rotational degrees of freedom at each control point. The performed numerical analyses demonstrate the effectiveness of the proposed HOM model implementation to ensure optimal convergence in both elastic and softening regime. Furthermore, the proposed approach allows the natural formulation of a localization limiter able to prevent strain localization and spurious mesh sensitivity known to be pathological issues for typical local strain-softening constitutive equations.      
}
\end{abstract}


\section{Introduction}
The nonlinear and fracturing behavior of quasi-brittle, heterogeneous materials such as fiber reinforced composites, toughened ceramics, and cementitious composites, among many others, is governed by weak spots and defects of their internal structures which cause initiation of cracks and localization of damage. In order to model such behavior, one can use discrete particle models or continuum models. Each type of model has its own advantages and disadvantages. 

In discrete particle models, the internal structure of materials is directly simulated through its geometric approximation and accounts for the actual intrinsic randomness specific of the selected length scale. Particles replicate the effect of major heterogeneities and their interaction is formulated by contact algorithms or through inter-particle lattice struts whose constitutive behavior models specific physical mechanisms. As such, discrete models have the advantage that many microscopic effects can be accounted for in a straightforward manner. However, they often lead to large computational systems, especially to simulate real structures, and require a significant, often overwhelming, amount of computational resources.

Continuum models are computationally less expensive. But most of the continuum models are based on the Cauchy continuum which cannot take into account features of the internal structure and does not possess a characteristic length. These type of models have some well-known drawbacks. First of all, they are not able to produce a mathematically well-posed problem in the softening regime causing strong mesh sensitivity in the numerical results \cite{bazant1988nonlocal, bazant1984continuum}. Secondly,  the dispersive nature of wave propagation, typical of many heterogeneous  materials,  can not be reproduced. Furthermore conventional continuum formulations predict unrealistic stress singularities at the tip of sharp cracks and fail to simulate size effect phenomena correctly \cite{rodriguez2011general,chang2002higher}.

In order to overcome these drawbacks, non-local continuum theories have been proposed in the literature. A popular approach consists in enriching continua with high order strain and stresses. Cosserat brothers' study \cite{cosserat1909theorie} was one of the initial attempt in this direction. They considered independent rotational degrees of freedom, modeling smaller scale kinematics, in addition to translational degrees of freedom. Later, the couple-stress theory was developed by Toupin \cite{toupin1962elastic}. In this model the only independent field is the displacement field that also determines the rotations. Eringen developed the so-called micropolar continuum theory \cite{eringen1965linear} and non-local elasticity \cite{eringen1992vistas}. Another type of high-order theory incorporate higher order gradients of displacements, mostly second gradient (strain gradient) and sometimes higher, into the constitutive model \cite{toupin1962elastic}. Mindlin \cite{mindlin1964micro,mindlin1965second,mindlin1968first} proposed an enhanced  elastic theory with microstructural effects by considering the potential energy density as a quadratic form of gradient of strain in addition to quadratic form of strains.  Aifantis and coworkers \cite{altan1992structure,aifantis1992role,altan1997some,aifantis1999strain,aifantis2003update} introduced a simplified gradient elasticity theory with one internal length scale which incorporates Laplacian of strain to enrich the continuum. They applied it for investigation of crack tips, localization of shear bands and to simulate size effect in torsion and bending. Various forms of gradient elasticity and their performance in static and dynamic applications are discussed by Askes and Aifantis \cite{askes2011gradient}. They also provide a procedure for the identification of the relevant characteristic lengths. In this work, the finite element implementation of gradient elasticity was also discussed. 

Chang et. al \cite{chang2002higher} developed a high-order gradient model based on a discrete particle structure, which includes both high-order strains and stresses. They investigated the possibility of formulating high-order constitutive equations with and without high order stresses. They found that, contrarily to the model with high-order stresses, the one without is unstable. Misra and Yang \cite{misra2010micromechanical} developed a model for cohesive materials based on microstructural concepts. They considered a representative volume of material as a collection of grains interacting through inter-granular force displacement relations. Inter-granular force displacement relations are formulated based on atomistic-level particle interactions. These force displacement relationships are then used to derive the incremental stress-strain relationship of a representative volume of material. Later Yang et. al \cite{yang2011higher} applied this type of microstructural granular based higher-order continuum theory to model the failure behavior of nanophased ceramics. They compared results obtained from the ab initio simulations with the result of higher-order continuum theory with good agreement.

Nonlinear nonlocal constitutive equations were also proposed without enrichment of the elastic part and by using integral or gradient approaches to enrich plasticity and damage theories. Integral type models are based replacing a local variable with its weighted spatial averaging over a certain volume of material. Pijaudier-Cabot and Ba\v zant  \cite{pijaudier1987nonlocal} pioneered research in this area. On the contrary, gradient plasticity or gradient damage models are based on enriching state variables with high-order terms.  A phenomenological strain gradient plasticity theory was developed by Fleck and Hutchinson \cite{fleck1997strain} in the context of couple-stress theory where only rotational part of strain gradients is taken into account. They extended the classical J2 plasticity to account for strain gradient effects. Later they reformulated their model and included also stretch gradients in addition to rotational ones \cite{fleck2001reformulation}. They investigated their model performance by applying it to the simulation of size effect on torsion of wires, sheet bending, indentation and void growth. Later, the so-called mechanism based theory of strain gradient plasticity (MSG) \cite{gao1999mechanism,Huang200099} was proposed on the basis of a multiscale framework linking statistically stored and geometrically necessary dislocations to plastic strains and strain gradient. A reformulation of mechanism-based strain gradient plasticity was given by \cite{yun2005reformulation} that involves the third-order tensor of higher-order stress to a much simpler version within Fleck and Hutchinson's theoretical framework of strain gradient plasticity theory. Huang et al. \cite{ Huang2004753} provide a conventional theory of mechanism-based strain gradient plasticity where the plastic strain gradient appears only in the constitutive model, and the equilibrium equations and boundary conditions are the same as the conventional continuum models. Jiang et al. \cite{jiang2001fracture} studied on fracture employing mechanism-based strain gradient plasticity.

Finite element implementation of material models that includes second gradient of displacements as degrees of freedom, require $C^1$ continuity at the inter-element boundaries or mixed element formulation in which both kinematic and static variables are defined as degrees of freedom.  Papanicolopulos et al. \cite{papanicolopulos2009three}, Zervos et al. \cite{zervos2001finite} developed $C^1$ type finite element for gradient elasticity. Fischear et al. \cite{fischer2011isogeometric} implemented isogeometric analysis for gradient elasticity in two dimensions. On the contrary, in mixed formulations, $C^0$ continuity element can be used but in this case first derivatives should be interpolated in addition to displacements which leads to formulations with a large number of degrees of freedom. These types of elements can be non-conforming and are required to fulfill the patch test \cite{zervos2009two,amanatidou2002mixed,phunpeng2015mixed}. Furthermore, Ru and Aifantis \cite{ru1993simple} developed a strategy based on operator split method for their simplified strain gradient model, in which the fourth-order partial differential equations of gradient elasticity are split into two separate second-order partial differential equations and they implement it using $C^0$ finite element.

\section{Higher Order Microplane Model}\label{sec:microplane}
Classical microplane model formulations, pionered by Ba\v{z}ant and coworkers
\cite{Bavzant1983Microplane,Bavzant1985Microplane}, assume that nonlinear phenomena such as, but not limited to, fracture, shearing, and plastic deformations occur on specific orientations within the internal structure of the material. Therefore, microplane models simulate these phenomena through vectorial constitutive laws formulated on the so called ``microplanes'' representing generic orientations in space. The strain vector on a generic microplane is obtained by projecting the strain tensor onto a local system of reference and the stress tensor is computed from tractions on all possible orientations in space through an energetic equivalence.

The initial formulations of microplane models focused on concrete materials and the most recent one, labeled as M7 \cite{caner2013microplane}, has demonstrated a remarkable ability to reproduce typical experimental data relevant to a large variety of loading conditions, from tension to confined compression; from quasi-static to highly dynamic conditions. Microplane models have also been successfully developed for other materials such as rock \cite{Bavzant2003Microplane}; rigid foam and shape memory alloys \cite{brocca2001microplane,brocca2002three};
fiber reinforced concrete \cite{beghini2007microplane}; as well as composite laminates \cite{Cusatis2007Spectral,beghini2007microplane,salviato2015spectral}. 

The aforementioned microplane formulations are
``local'' in the sense that the stress tensor at a given point is only
function of the strain tensor at the same point.
Unfortunately, such  constitutive equations suffer from 
mesh sensitivity and spurious energy dissipation \cite{Bavzant1983crack} when
softening behavior is simulated. To overcome this problem, nonlocal microplane models were proposed by various authors \cite{Bavzant1990nonlocal,Bavzant2004nonlocal,DiLuzio2007} by exploiting integral methods, in which the nonlinear behavior at one point depends on a weighted average of the strain in the neighborhood of that point. These approaches while preventing mesh sensitivity quite effectively lead to a significant increase of the computational cost of simulations.

An alternative to integral microplane models is to adopt strain-gradient formulations such as the ones proposed by Kuhl et Al. \cite{kuhl1999simulation,kuhl2000anisotropic} and, more recently, by Cusatis and Zhou \cite{cusatis2013high}. In the latter, starting from a discrete particle model, they derived a high order microplane (HOM) theory which includes Cosserat theory and strain gradients elasticity as a special cases. The HOM model possess two characteristic lengths: the first associated with the spacing of flaws in the material internal structure and related to the gradient character of the continuum; and the second associated with the size of these flaws and related to the micro-polar character of the continuum. 

Following Cusatis and Zhou \cite{cusatis2013high}, one can calculate microplane strains and curvatures as 
\begin{equation}
\label{eq:def_microplane_strain}
\varepsilon_{\beta}= P_{ij}^{\beta} \gamma_{ij} +r_0 \mathcal{P}_{ijk}^{\beta} \Gamma_{ijk}; ~~~ \chi_{\beta}=P_{ij}^{\beta} \kappa_{ij}; ~~~ \beta=N,M,L
\end{equation}
where $\gamma_{ij}=u_{j,i}-\epsilon_{ijk} \varphi_k$ is the strain tensor, $\kappa_{ij}=\varphi_{j,i}$ is the curvature tensor; $\Gamma_{ijk}=\gamma_{ij,k}$  is the strain gradient tensor also referred to as high order strain tensor; $u_i$, $\varphi_i$ are displacement and rotation, respectively, at a generic position $x_i$ ($i=1,2,3$) in a 3D continuum. The second order tensor $P_{ij}^{\beta}=e_i^N e_j^{\beta}$ and the third order tensor $\mathcal{P}_{ijk}=e_i^N e_j^{\beta} e_k^N$ denote projection operators for strain and strain gradient, respectively, in which  $e_i^{\beta}  \left( \beta=N,M,L \right)$ are unit vectors defining a local system of reference on each microplane (see Fig~\ref{geometry}). The characteristic length $r_0$, assumed to be a material property, can be interpreted to be the average half spacing of weak spots in the material internal structure\cite{cusatis2013high}.

By using appropriate vectorial constitutive equations, the microplane stresses and microplane couple stresses can be calculated from the microplane strains and microplane curvatures: formally, one can write $\sigma_\beta=\mathcal{F}_\beta(\epsilon_N, ..., \chi_N, ...)$ and $\mu_\beta=\mathcal{G}_\beta(\epsilon_N, ..., \chi_N, ...)$, which can be used to obtain stress, couple stress, and high order stress tensors as

\begin{equation}\label{eq:stress_integration_sphere}
\sigma_{ij}=\frac{3}{4 \pi} \int_{\Gamma_1} \sigma_{\beta} P_{ij}^{\beta} \mathrm{d}S ; ~~~ \mu_{ij}=\frac{3}{4 \pi} \int_{\Gamma_1} \mu_{\beta} P_{ij}^{\beta} \mathrm{d}S ; ~~~
\Sigma_{ijk}=\frac{3r_0}{4 \pi} \int_{\Gamma_1} \sigma_{\beta} \mathcal{P}_{ijk}^{\beta} \mathrm{d}S 
\end{equation}
where the integrals are calculated over the surface of a unit sphere, $\Gamma_1$; $\beta=N,M,L$; and summation rule applies over $\beta$. 

Furthermore, stress, couple stress, and high order stress tensors must satisfy the conservation of linear and angular momenta which reads
\begin{equation}
\label{eq:equilibrium}
\pi_{ji,j}+b_i=0; \quad \mu_{ji,j}+\epsilon_{ijk} \pi_{jk}=0
\end{equation}
where $\pi_{ij}=\sigma_{ij}-\Sigma_{ijk,k}$ is the so-called effective stress tensor; $b_i$ denote body forces and moments; and $\epsilon_{ijk}$ is the Levi-Civita permutation symbol.

The weak form  of equilibrium that is used for the finite elelemnt formulation read
\begin{equation} \label{Eq:Pricipal-Virtual-Work}
\begin{split}
\int_\Omega \left( \sigma_{ij} \delta \gamma_{ij} + \Sigma_{ijk} \delta \Gamma_{ijk} + \mu_{ij} \delta \kappa_{ij} \right) \textrm{d} V = 
\int_\Omega b_i \delta u_i 
\textrm{d} V + \\  
+ \int_{ \partial \Omega } \left( m_i \delta \varphi_i + t_i \delta
  u_i + h_i D\delta u_i \right) \textrm{d} S
+ \oint_C c_i\delta u_i \textrm{d} L
\end{split}
\end{equation}

Finally, the differential boundary value problem presented above must be completed with appropriate boundary conditions. With reference to Fig. \ref{HOM_BC}, one can write, on the volume boundary, $\pi_{ji} n_j -\mathrm{D}_j \left( \Sigma_{ijk} n_k \right) + \left( \mathrm{D}_l n_l \right) h_i = \bar{t}_i$ or $u_i=\bar{u}_i$, $\Sigma_{ijk} n_j n_k = \bar{h}_i$ or $\mathrm{D}u_i = \mathrm{D}\bar{u}_i$, $\left( \mu_{ji} - \epsilon_{pki} \Sigma_{pkj} \right) n_j =\bar{m}_i$ or $\varphi_i=\bar{\varphi}_i$, and $ \llbracket  r_i n_k \Sigma_{pkj}  \rrbracket= \bar{c}_i$, where $n_i$ is the unit vector orthogonal to the boundary, $\mathrm{D}_i \equiv (\delta_{ij}-n_in_j)\partial/\partial x_j$ is the surface gradient differential operator, $\delta_{ij}$ = Kronecker delta,  $\bar{t}_i$=applied tractions,  $\bar{u_i}$=applied displacements, $\bar{h_i}$=applied moment energetically conjugate to $\mathrm{D}u_i $, $\mathrm{D}u_i=u_{i,j}n_j$ is  the displacement normal derivative, $\mathrm{D}\bar{u}_i $ = applied displacement normal derivative, $\bar{m}_i$=applied moments energetically conjugate to rotations,  $\bar{\varphi_i}$=applied rotations, $\bar{c}_i$ = applied edge traction acting along a sharp edge, $\llbracket \cdot \rrbracket$ is the jump operator defined on a sharp edge, $r_j=\epsilon_{pqj} s_p n_q$, and $s_i$ is the unit vector tangent to a sharp edge (See Fig. \ref{HOM_BC}). Detailed derivation of Eqs. \ref{eq:equilibrium} and boundary conditions can be found in Cusatis and Zhou \cite{cusatis2013high}.

\section{Isogeometric finite element implementation}
\subsection{Isogeometric Description of Geometry}
Isogeometric analysis (IGA), introduced by Hughes et al. \cite{hughes2005isogeometric}, combines computer aided design (CAD) and finite element method (FEM) technologies. In IGA, the basis functions used in CAD to describe exactly the geometry of a certain volume of material are also used to discretize field variables describing, for example, its deformation and mechanical behavior. CAD technologies commonly employ Non-Uniform Rational B-Splines (NURBS) basis functions which are built from B-splines \cite{farin1995nurb}. B-splines are piecewise polynomials composed of a linear combination of basis functions defined recursively as
\begin{equation}
\label{eq:basisP}
N_{i}^p(\xi)=\frac{\xi-\xi_i} {\xi_{i+p}-\xi_i} N_{i}^{p-1}(\xi)
+ \frac{ \xi_{i+p+1}-\xi} {\xi_{i+p+1}-\xi_{i+1} } N_{i+1}^{p-1}(\xi)
\end{equation}
where $i=1,..., n$; $n$ and $p$ are number and maximum order, respectively, of the basis functions. The  \emph{knots vector}, $\mathbf{X}=\{\xi_1,\xi_2, ...,\xi_{n+p+1}\}$ is a set of non-decreasing real numbers called \emph{knots} and representing coordinates in the parameter domain. If all knots are equally spaced, the knot vector is called uniform, otherwise they are non-uniform. For $p=0$, $N_{i}^0(\xi)=1$ if $\xi$ is contained in the interval $[\xi_i,\xi_{i+1}]$, and $N_{i}^0(\xi)=0$ otherwise. For $p = 1$ B-spline basis functions are identical to basis function of classical constant-strain finite element formulations. B-spline basis functions are linearly independent, have partition of unity property and their support is compact. However, they, in general, do not satisfy Kronecker delta property unless coincident knots exists in $\mathbf{X}$. A B-spline is said to be open if its first and last knots appear $p+1$ times. Open knot vectors are interpolatory at the end knots which means they satisfies Kronecker delta property at those knots \cite{nguyen2015isogeometric}.
 
B-splines cannot represent some common geometrical shapes, such as circular and ellipsoidal sections, exactly. This can be achieved by using Non-Uniform Rational B-Splines (NURBS), which can be defined in terms of B-spline basis functions as follows:
\begin{equation}
\label{eq:NurbasisP}
R_{i}^{p}(\xi)=\frac{N_{i}^{p}(\xi)} {\sum_{j=1}^n N_{j}^{p}(\xi) w_j}
\end{equation}
where $w_j$ are weighting factors that are selected depending upon the type of curve to be represented exactly \cite{hughes2005isogeometric, nguyen2015isogeometric}. 

A NURBS curve can in general defined as linear combinations of the the functions defined in Eq. \ref{eq:NurbasisP}. Hence, in 3D, the generic position vector $\mathbf{x}=[x_1~x_2~x_3]^T$ can be expressed as

\begin{equation}
\label{eq:nurbsolid}
\mathbf{x}(\xi,\eta,\zeta)=\sum_{ i=1}^{n} \sum_{j=1}^{m}  \sum_{k=1}^{l} R_{ijk}^{pqr}(\xi,\eta,\zeta) \mathbf{x}_{ijk}=\sum_{ I=1}^{N_{cp}} R_{I}^{pqr}(\xi,\eta,\zeta) \mathbf{x}_{I}
\end{equation}
where 
\begin{equation}
\label{eq:nurbsolid2}
R_{I}^{pqr}(\xi,\eta, \zeta)=R_{ijk}^{pqr}(\xi,\eta, \zeta)=\frac{N_{i}^{p}(\xi) M_{j}^{q}(\eta) P_{k}^{r}(\zeta) w_{ijk}} {\sum_{\hat{i}=1}^{n} \sum_{\hat{j}=1}^{m} \sum_{\hat{k}=1}^{l} N_{\hat{i}}^{p}(\xi) M_{\hat{j}}^{q}(\eta) P_{\hat{k}}^{r}(\zeta) w_{\hat{i}\hat{j}\hat{k}}} 
\end{equation}
where $N_i^p$, $M_j^q$, $P_k^r$ are the basis function defined according to Eq. \ref{eq:basisP} and with reference to given knot vectors for each direction $\mathbf{X}=\{\xi_1,\xi_2, ...,\xi_{n+p+1}\}$ ,  $\mathbf{Y}=\{\eta_1,\eta_2, ...,\eta_{m+q+1}\}$ and $\mathbf{Z}=\{\zeta_1,\zeta_2, ...,\zeta_{k+r+1}\}$; $\mathbf{x}_{I}\equiv\mathbf{x}_{ijk}$ is the position vector of a \emph{control point} $I$ (a.k.a. \emph{node}, per the usual computational mechanics terminology) defined on a 3D grid ($i=1,..,n$, $j=1,..,m$, $k=1,..,l$) associated to the knot vectors $\mathbf{X}$, $\mathbf{Y}$, and $\mathbf{Z}$;  $N_{cp}=m \times n \times l$ is the total number of control points.

\subsection{Implementation of isogeometric element}
In this study a isogeometric finite element is implemented in 3D with interpolation of displacement and rotational fields assumed to be independent. This element features 6 degrees of freedom (3 displacement and 3 rotations) at each control point.  Following the isoparametric concept, the displacement field $\boldsymbol{u}$ and rotation field $\boldsymbol{\varphi}$ are formulated using the same NURBS basis functions of the geometry and, for a given element, they can be approximated in terms of element's control point displacements and rotations as follows

\begin{equation}
\label{eq:displacement_field}
\mathbf{u}=\displaystyle\sum_{ I=1}^{N_{cp}} R_{I} \mathbf{u}_I^e \quad \Phi=\displaystyle\sum_{ I=1}^{N_{cp}} R_{I} \boldsymbol{\varphi}_I^e
\end{equation}
where $N_{cp}$ is the number of control points that are supported by one element and $\mathbf{u}^e=\left[ u_{xi}, u_{yi}, u_{zi} \right]$ denotes the nodal displacement vector and $ \boldsymbol{\varphi}^e=\left[ \varphi_{xi}, \varphi_{yi}, \varphi_{zi} \right] $ the nodal rotation vector. It is worth nothing that in this study the adopted shape functions in Eq.~\ref{eq:displacement_field} are quadratic which can be obtained from Eq.~\ref{eq:nurbsolid2} with $p=q=r=2$ and $N_{cp}=27$ control points per element. In Eq.~\ref{eq:displacement_field} and throughout the rest of the paper, the shape function superscripts are dropped for clarity of notation. Low and high order strain fields are obtained computing the spatial derivatives of the displacement field:
\begin{equation}
\label{eq:strain_vec}
\mathbf{e}=\mathbf{L} \mathbf{d}^e= \mathbf{LR} \mathbf{d}^e= \mathbf{B} \mathbf{d}^e
\end{equation}  
where $\mathbf{e}=\left[ \gamma_{ij}, \kappa_{ij}, \Gamma_{ijk} \right]^T$ is the strain vector collecting all the strains, curvature and high order strain components (45 in total),  $\mathbf{d}$ is the displacement vector, $\mathbf{d}=\left[ u_i, \phi_i \right]$ collecting all degrees of freedom of the element ($6 \times Ncp$ in total), $\mathbf{L}$ is a differential operator, $\mathbf{R}$ is the shape function matrix and $\mathbf{B}$ is the strain-displacement matrix obtained by applying the differential operator $\mathbf{L}$ to the shape functions. The $\mathbf{B}$ matrix can be subdivided into sub-matrices for low-order and high-order terms and the separate contribution of different nodes:
\begin{equation}
\label{eq:B_matrix}
\mathbf{B}=
\begin{bmatrix}
\mathbf{B}_1 & \mathbf{B}_2    & ...          & \mathbf{B}_{N_{cp}}
\end{bmatrix} \quad ; \quad
\mathbf{B_I}=
\begin{bmatrix}
\mathbf{B}_{I \gamma} &\mathbf{B}_{I \kappa} &	\mathbf{B}_{I \Gamma}  			
\end{bmatrix}^T
\end{equation}
in which
	\begin{equation}
	\small	
\mathbf{B}_{I \gamma}=
	\left[
	\begin{array}{cccccc}
	R_{I,x} & 0 & 0 & 0 & 0 & 0 \\ 
	0 & R_{I,y} & 0 & 0 & 0 & 0 \\ 
	0 & 0 & R_{I,z} & 0 & 0 & 0 \\ 
	0 & R_{I,x} & 0 & 0 & 0 & -1 \\ 
	0 & 0 & R_{I,y} & -1 & 0 & 0 \\ 
	R_{I,z} & 0 & 0 & 0 & -1 & 0 \\ 
	R_{I,y} & 0 & 0 & 0 & 0 & 1 \\ 
	0 & R_{I,z} & 0 & 1 & 0 & 0 \\ 
	0 & 0 & R_{I,x} & 0 & 1 & 0 \\ 	
	\end{array} 
	\right],
	\qquad
	\mathbf{B}_{I \kappa}	=
	\left[
	\begin{array}{cccccc}	
	0 & 0 & 0 & R_{I,x} & 0 & 0 \\ 
	0 & 0 & 0 & 0 & R_{I,y} & 0 \\ 
	0 & 0 & 0 & 0 & 0 & R_{I,z} \\ 
	0 & 0 & 0 & 0 & R_{I,x} & 0 \\ 
	0 & 0 & 0 & 0 & 0 & R_{I,y} \\ 
	0 & 0 & 0 & R_{I,z} & 0 & 0 \\ 
	0 & 0 & 0 & R_{I,y} & 0 & 0 \\ 
	0 & 0 & 0 & 0 & R_{I,z} & 0 \\ 
	0 & 0 & 0 & 0 & 0 & R_{I,x}
	\end{array} 
	\right]
	\end{equation}

\begin{equation}
  \scriptsize
  \resizebox{\textwidth}{!}{%
$	\mathbf{B}_{I \Gamma}=
	\left[
	\arraycolsep=0.0pt
	\begin{array}{ccccccccccccccccccccccccccc}	
R_{I,xx} & 0 &0 &0 &0 &R_{I,zx} &R_{I,yx} &0 &0 &R_{I,xy} &0 &0 &0 &0 &R_{I,zy} &R_{I,yy} &0 &0 &R_{I,xz} &0 &0 &0 &0 &R_{I,zz} &R_{I,yz} &0 &0 \\
0 &R_{I,yx} &0 &R_{I,xx} &0 &0 &0 &R_{I,zx} &0 &0 &R_{I,yy} &0 &R_{I,xy} &0 &0 &0 &R_{I,zy} &0 &0 &R_{I,yz} &0 &R_{I,xz} &0 &0 &0 &R_{I,zz} &0\\
0 &0 &R_{I,zx} &0 &R_{I,yx} &0 &0 &0 &R_{I,xx} &0 &0 &R_{I,zy} &0 &R_{I,yy} &0 &0 &0 &R_{I,xy} &0 &0 &R_{I,zz} &0 &R_{I,yz} &0 &0 &0 &R_{I,xz}\\
0 &0 &0 &0 &-R_{I,x} &0 &0 &R_{I,x} &0 &0 &0 &0 &0 &-R_{I,y} &0 &0 &R_{I,y} &0 &0 &0 &0 &0 &-R_{I,z} &0 &0 &R_{I,z} &0 \\
0 &0 &0 &0 &0 &-R_{I,x} &0 &0 &R_{I,x} &0 &0 &0 &0 &0 &-R_{I,y} &0 &0 &R_{I,y} &0 &0 &0 &0 &0 &-R_{I,z} &0 &0 &R_{I,z}\\
0 &0 &0 &-R_{I,x} &0 &0 &R_{I,x} &0 &0 &0 &0 &0 &-R_{I,y} &0 &0 &R_{I,y} &0 &0 &0 &0 &0 &-R_{I,z} &0 &0 &R_{I,z} &0 &0\\
	\end{array} 
	\right]^T $ %
	}		
	\end{equation}
Once the strain displacement matrix, $\mathbf{B}$, is obtained the element stiffness matrix and the internal force vector can be calculated easily by employing appropriate quadrature rules:
\begin{equation}
\label{eq:stiffnessmat}
    \mathbf{K}=\begin{bmatrix}
    \mathbf{K}_{11} 	& \mathbf{K}_{12}  			& \cdots    		\\
    \mathbf{K}_{21}				& \mathbf{K}_{22} &      			\\
     \vdots				& 			& 		\ddots							 \\				
    \end{bmatrix}, \quad 
    \mathbf{f}^T=\left[ \mathbf{f}_1^T, \mathbf{f}_2^T ...    \right], \quad 
	\mathbf{K}_{IJ}=\int_{\Omega} \mathbf{B}_I^T \mathbf{D} \mathbf{B}_J \mathrm{d}\Omega, \quad  	\mathbf{f}_I=\int_{\Omega} \mathbf{B}_I^T 	\mathbf{s} \mathrm{d}\Omega
\end{equation}
where $\mathbf{s}=\left[\sigma_{ij},\mu_{ij},\Sigma_{ijk}	\right]^T$ is the stress vector collecting all components of the stress tensor, couple stress tensor and high order stress tensor; and $\mathbf{D}$ is the stiffness matrix, $\mathbf{s}=\mathbf{D} \mathbf{e}$. Even though there are recent studies attempting to derive optimal quadrature rules for IGA \cite{auricchio2012simple,schillinger2014reduced,hiemstra2016optimal}, in this study,  Gauss quadrature is used and number of Gauss points in each direction is assumed equal to be 1 plus the order of basis function in each direction, that is three for quadratic element which leads to 27 Gauss points per element. Gauss quadrature was originally developed for polynomial function, but it has been employed successfully for IGA element as well \cite{hughes2005isogeometric,nguyen2015isogeometric}.

It is important to point out that the proposed element formulation implicitly assumes that the high order surface moments $\mathbf{h}$ conjugate to the normal gradient of the displacement field (see Sec. \ref{sec:microplane} and Fig. \ref{HOM_BC}) are zero. This is due to the fact that only displacement and rotation degrees of freedom are introduced and no kinematic boundary conditions can be imposed on the normal gradient of displacements. 

\section{Numerical Results}
The performance of the proposed isogeometric implementation of the high order microplane model is demonstrated in this section with reference to classical examples for gradient elasticity, Cosserat elasticity and strain softening. 
  
\subsection{Strain gradient elasticity solution of cantilever beam problems}
The exact analytical solution \cite{papargyri2003bending} for a cantilever beam subjected to a concentrated load at the free end is used here as a reference for comparison with the strain gradient version of the high order microplane model. To obtain the macroscopic constitutive equations used to derive the exact solution, the microplane constitutive equation for gradient elasticity are formulated as follows: 
\begin{equation}
\sigma_N=E_V \varepsilon_V + E_D \varepsilon_D+E_N^{\Gamma} \epsilon_N^{\Gamma}; ~~~\sigma_M=E_T  \varepsilon_M^{\gamma} +E_T^{\Gamma} \varepsilon_M^{\Gamma}; ~~~ \sigma_L=E_T  \varepsilon_L^{\gamma} +E_T^{\Gamma} \varepsilon_L^{\Gamma}
\label{HOM:strain-gradient-elasticity}
\end{equation}
and $\mu_N=\mu_M=\mu_L=0$, where $\varepsilon_{\beta}=\varepsilon_{\beta}^{\gamma}+\varepsilon_{\beta}^{\Gamma};~~~ 
\varepsilon_{\beta}^{\gamma}=P_{ij}^{\beta} \gamma_{ij};~~~ 
\varepsilon_{\beta}^{\Gamma}=r_0 \mathcal{P}_{ijk}^{\beta} \Gamma_{ijk}, ~~~ \left(\beta=N,M,L\right) $,  $\varepsilon_V=V_{ij}\gamma_{ij}$, $V_{ij}=\delta_{ij}/3$, $\delta_{ij}$ is the Kronecker delta, $\varepsilon_D^{\gamma}=\varepsilon_N^{\gamma}-\varepsilon_V$, and $E_V$, $E_D$, $E_T$ are volumetric, deviatoric and shear elastic moduli, respectively. As already shown by Cusatis and Zhou [Ref], Eq. \ref{HOM:strain-gradient-elasticity} lead to classical elastic constitutive equations if $r_0=0$, $E_V=E/(1-2\nu)$ and $E_D=E_T=E/(1+\nu)$ where $E=$ Young's modulus and $\nu=$ Poisson's ratio.
With the formulation in Eq.~\ref{HOM:strain-gradient-elasticity}, the microplane integration (see Eq.~\ref{eq:stress_integration_sphere}) can be done analytically and explicit form of stress, couple stress, and high order stress tensor can be obtained as follows: 
\begin{gather}
\sigma_{ij}= E_V \epsilon_V V_{ij} + \left[ E_D A_{ijkl}+E_T \left(B_{ijkl}+C_{ijkl}\right) \right] \gamma_{kl}, ~~ \mu_{ij}=0 \nonumber \\
\Sigma_{ijk}=r_0^2 \left[ E_V D_{ijklmp} + E_T \left(E_{ijklmp}+F_{ijklmp}\right) \right] \Gamma_{lmp}
\end{gather}  
where $A_{ijkl}=3 \int_{\Gamma_1}  P_{ij}^{D} P_{kl}^{D} ds / 4\pi$, $B_{ijkl}=3 \int_{\Gamma_1}  P_{ij}^{M} P_{kl}^{M} ds / 4 \pi$,  $C_{ijkl}=3 \int_{\Gamma_1}  P_{ij}^{L} P_{kl}^{L} ds/ 4\pi$,  $P_{ij}^D=P_{ij}^N-V_{ij}$,  $D_{ijklmp}=3 \int_{\Gamma_1}  \mathcal{P}_{ijk}^{N} \mathcal{P}_{lmp}^{N} ds/ 4\pi$, $E_{ijklmp}=3 \int_{\Gamma_1}  \mathcal{P}_{ijk}^{M} \mathcal{P}_{lmp}^{M} ds/ 4 \pi$ and $F_{ijklmp}=3 \int_{\Gamma_1}  \mathcal{P}_{ijk}^{L} \mathcal{P}_{lmp}^{L} ds/ 4 \pi$.

 By assuming $E_D=E_T=E_0/(1+ \nu)$, $E_V=E_0/(1-2 \nu)$, $E_T^{\Gamma}=0$ and $E_N^{\Gamma}=E$ then the gradient elasticity formulation in Beskou et al \cite{papargyri2003bending} is obtained for the components $\sigma_{11}$ and $\Sigma_{111}$. It must be observed however that this microplane formulation leads to other non-zero high order stress components which are absent in the analytical solution.

 A cantilever beam, 1000 mm long with $100 \times 25$ mm cross-section, is modeled with quadratic isogeometric solid elements. The beam is discretized with 4 different meshes with $10 \times 1 \times 1$, $20 \times 2 \times 1$, $40 \times 4 \times 1$ and $80 \times 8 \times 2$ elements, respectively. Essential boundary conditions are applied at both ends of the beam. E=25000  MPa, $\nu$=0.2, and $2 r_0=0.2L$ are assumed in the calculations. In Fig.~\ref{fig:cantilever_beam_con} the relative displacement (normalized with displacement value obtained from the Cauchy solution) as a function of position along the beam is given for different meshes and compared with analytical result given by Beskou et al \cite{papargyri2003bending}. It can be seen from the figure that with mesh refinement the tip displacement converges to a certain value. The overall calculated elastic curve is in general good agreement with the analytical solution, to which, however, it does not converge exactly. The main reason for this discrepancy is that in the analytical solution, they only consider one high-order term $\Gamma_{xxx}=\gamma_{xx,x}$, but the proposed HOM model is a 3D model and it features automatically other high order terms.  

Fig.~\ref{fig:cantilever_beam_disp} reports the obtained response for different values of the internal length scale, $r_0$. As expected, with the increase of the internal length scale, the beam response becomes stiffer. Also, internal length that tends to zero, the response converges to the classical Bernoulli beam solution.

The second beam example is relevant to an experimental study carried out by Lam et al. \cite{lam2003experiments} on epoxy beams. Tests were carried out in plane strain conditions on end-point loaded cantilever beams with 20, 38, 75, 115 $\mu$m depth and length-to-depth ratio equal to 10. The elastic properties of the epoxy are reported as $E=1.44$ GPa and $\nu =0.38$ \cite{lam2003experiments}. The microplane simulations were performed with $E_T^{\Gamma}=E_T$ and $E_N^{\Gamma}=E_V$ and the material characteristic length was optimized according to the experimental result for the 20 $\mu$m beam. The best fit is obtained with $2r_0=33.2$ $\mu$m. Based on these material parameters, the other beams are simulated. 3D quadratic isogeometric elements are used and the beams are discretized with $40 \times 4 \times 1$ elements. In order to provide plane strain condition, the lateral displacement ($u_z$) of nodes on external surfaces (x-y plane) are restricted. Figure~\ref{fig:cantilever_beam_stiffness} illustrates the change of normalized bending stiffness, $D=P L^3 /\left(3 w b h^3 \right)$, predicted by the proposed model and obtained in the experiments. In the definition of $D$, $P$ is the applied load; $w$ is the corresponding displacement; $b$,$h$ and $L$ are beam thickness, depth and length, respectively. The model can reproduce very well the beam behavior exhibiting size effect: the smaller the thickness the larger the normalized bending stiffness \cite{lam2003experiments}. This behavior can not be produced by classical beam, which always gives constant value for bending stiffness corresponding to the current formulation for large thickness values.    

\subsection{Cosserat Solution of a Stress Concentration Problem}
The presented high order microplane theory reduced to a Cosserat continuum when  the material characteristic length, $r_0$, is equal to zero, and the microplane constitutive equations are formulated as follows:  
\begin{gather} 
 \sigma_N=E_V \varepsilon_V + E_D \varepsilon_D; ~~~\sigma_M=E_T  \varepsilon_M; ~~~ \sigma_L=E_T  \varepsilon_L \\
\mu_N=W_V \chi_V + W_D \chi_D; ~~~\mu_M=W_T  \chi_M; ~~~ \chi_L=W_T  \chi_L
\label{HOM:cosserat-elasticity}
\end{gather}
where $\varepsilon_V$ is the volumetric strain, $\varepsilon_D=\varepsilon_N-\varepsilon_V$, $\chi_V=V_{ij} \chi_{ij}$, $\chi_D=\chi_N-\chi_V$. Also in this case, the microplane integration ( Eq.~\ref{eq:stress_integration_sphere}) can be done analytically and explicit form of stress and couple stress tensors are obtained as follows:
\begin{gather}
\label{eq:cosserat_stress}
\sigma_{ij}= E_V \varepsilon_V V_{ij}  + \left[ E_D A_{ijkl}+E_T \left(B_{ijkl}+C_{ijkl}\right) \right] \gamma_{kl} \nonumber \\
\mu_{ij}= W_V \chi_V V_{ij} + \left[ W_D A_{ijkl}+W_T \left(B_{ijkl}+C_{ijkl}\right) \right] \kappa_{kl}
\end{gather}
where the tensors $V_{ij}$, $A_{ijkl}$, $B_{ijkl}$ and $C_{ijkl}$ are the same ones introduced in the previous section. Cusatis and Zhou \cite{cusatis2013high} derived the relationship between the microplane parameters and the ones of Cosserat elasticity as $E_V=3 \lambda + 2 \mu + \chi$, $E_D=5 \mu + \chi$, $E_T=\chi$ and $W_V=3\pi_1+\pi_2+\pi_3$, $W_D=\pi_2+4\pi_3$, $W_T=\pi_2-\pi_3$.

In order to demonstrate the efficiency of the isogeometric element for the integration of the Cosserat theory, this section discusses a well-known stress concentration problem consisting of an infinite plate with a circular hole subjected to a uniaxial far field tension. The Cosserat continuum has a characteristic length which arises from the unit mismatch between stresses and couple stresses. This effects the stress concentration factor around the hole. Analytical solutions of this problem for classical Cosserat continuum was given by Eringen \cite{eringen1967theory} and for Couple Stress theory by Mindlin \cite{mindlin1963influence}  in 2D case. In both references the stress concentration factor is defined as $SCF=\sigma_x(0,r)/\sigma_{x}^{\infty}= (3+F)/(1+F)$ where $F=[8 l_b^2/l_c^2 (1-\nu)] / [  4+(r/l_c)^2+2 (r/l_c) K_0(r/l_c)/K_1(r/l_c) ]$, $r$ is the radius of the hole, $l_b=\left[\pi_3 / 2 \left(2 \mu + \chi \right)\right]^{1/2} $, $l_c=\left[\left(\pi_2+\pi_3\right) /  \left(2 \mu + \chi \right)\right]^{1/2}$ \cite{pothier1994three} are the material characteristic lengths, and $K_0$ and $K_1$ are the second kind modified Bessel functions of order zero and one, respectively. Mindlin gives exactly the same equation for the case of $l_b^2/l_c^2=1$. Furthermore, it is convenient to define \cite{pothier1994three} the coupling factor, $N_c=\left[ \chi/2 \left( \mu+ \chi \right) \right]$ , which is a non-dimensional quantity taking values between 0 and 1. For $N_c=1$ the couple-stress theory is recovered and for $N_c=0$ the Cauchy continuum is obtained \cite{pothier1994three}.  

This example was also simulated by Pothier and Rencis \cite{pothier1994three} and Nakamura et. al. \cite{sachio1984finite}. In the finite element simulations discussed in this section, the material parameters are taken from Ref.~\cite{pothier1994three} and they are reported in Table~\ref{table_SCF_mat}.  Four simulations with different coupling factor are performed by assuming plate dimensions of $300 \times 300 \times 10$ mm and hole radius of $10$ mm. Considering the symmetry along the x- and y-axes, only one quarter of the specimen is modeled as shown in Figure~\ref{plate_hole}. Applied boundary conditions are also specified in Figure~\ref{plate_hole}. A 3D finite element model was set up with quadratic isogeometric solid element and the IGA meshes was obtained by using IGAFEM \cite{nguyen2015isogeometric}.  Two different meshes are used with 1024 ($32 \times 32$) and 4096 ($64 \times 64$) elements. For both mesh only one element is used along the thickness. The coarser mesh is shown in Fig. \ref{plate_hole} where a zoom-in highlights the geometry of one element and the corresponding control points. Two more layers of control points are present through the thickness (for a total of 27 control points) but they are not visible in the figure. Table~\ref{table_SCF} presents the comparison of the obtained stress concentration factors with the analytic solution. The results show very good agreement with analytical values with error less than 3.7 \% and 0.9 \% for mesh-1 and mesh-2, respectively.

\subsection{Localization Behavior of a Softening Bar in Tension}

This section presents the simulation of softening behavior with the objective of evaluating the effectiveness of the regularization strategy proposed by Cusatis and Zhou within the presented high order isogeometric implementation \cite{cusatis2013high}.

The adopted softening constitutive equation follows the formulation proposed by Cusatis et Al. \cite{Cusatis2011Lattice}. The microplane couple stresses are assumed to be zero $\mu_N=\mu_M=\mu_L$=0. The microplane stresses are computed with damage-like constitutive equations as follows
\begin{equation}
\label{eq:microplane_stress}
\sigma_N= \frac{\sigma}{\epsilon}\epsilon_N \quad
\sigma_M=\alpha \frac{\sigma}{\epsilon}\varepsilon_M \quad
\sigma_L=\alpha \frac{\sigma}{\epsilon}\varepsilon_L
\end{equation}
where $\alpha$ is a material parameter, $\sigma =(\sigma_N^2+\sigma^2_T/\alpha)^{1/2}$ is the effective stress, $\sigma_T=(\sigma_M^2+\sigma_L^2 )^{1/2}$ is the total shear stress, $\epsilon =(\epsilon_N^2+\alpha \epsilon_T^2)^{1/2}$ is the effective strain, and $\epsilon_T=(\epsilon_M^2+\epsilon_L^2)^{1/2}$ is the total shear strain. In the elastic regime, the effective stress is proportional to the effective strain: $\sigma = E_0 \epsilon$, in which $E_0$ is the microplane normal modulus. $E_0$ and $\alpha$ can be related to Young's modulus, $E$, and Poisson's ratio, $\nu$, through the following expressions: $E_0=E/(1-2\nu)$ and $\alpha= (1-4\nu) /(1+\nu)$. The nonlinear behavior is imposed by computing the effective stress incrementally, $\dot{\sigma}=E_0\dot{\epsilon}$, and satisfying the inequality $0\leq \sigma \leq \sigma _{bt} (\epsilon, \omega) $ where $\sigma_{bt} = \sigma_0(\omega) \exp \left[-H_0(\omega) \langle \epsilon-\epsilon_0(\omega) \rangle / \sigma_0(\omega)\right]$, $\langle x \rangle=\max \{x,0\}$, $\epsilon_0(\omega)=\sigma_0(\omega)/E_0$, and $\tan(\omega) =\epsilon_N / \sqrt{\alpha} \epsilon_{T}$ = $\sigma_N \sqrt{\alpha} / \sigma_{T}$. The post peak softening modulus is defined as $H_{0}(\omega)=H_{t}(2\omega/\pi)^{n_{t}}$, where $H_{t}$ is the softening modulus in pure tension ($\omega=\pi/2$) expressed as $H_{t}=2E_0/\left(\ell_t/\ell_0-1\right)$; $\ell_t=2E_0G_t/\sigma_t^2$; $\ell_0=2r_0$; and $G_t$ is the mesoscale fracture energy. In the previous equations, the effective strength is formulated in such a way to have a smooth transition between pure tension ($\omega=\pi/2$) and pure shear ($\omega=0$) with parabolic variation in the $\sigma_N$ versus $\sigma_T/\sqrt{\alpha}$ plane: $\sigma_{0}(\omega )=\sigma _{t}r_{st}^2\Big(-\sin(\omega) + \sqrt{\sin^2(\omega)+4 \alpha \cos^2(\omega) / r_{st}^2}\Big) / [2 \alpha \cos^2(\omega)]$, where $r_{st} = \sigma_s/\sigma_t$ is the ratio of shear strength to tensile strength.

By assuming $E=25$ GPa, $\nu=0.2$, $r_0=5$ mm, $\sigma_t=$3 MPa, $\sigma_s/\sigma_t=$4 MPa, and $\ell_t=$100 mm, a prismatic bar subject to uniaxial tension is modeled with 3D quadratic isogeometric solid elements. The bar is 100 mm long and has $75 \times 75$ mm rectangular cross section. One end of the bar is fixed in all directions and a displacement is prescribed at opposite end in the longitudinal direction. Four different meshes are used with $10\times1\times1$, $20\times1\times1$, $40\times1\times1$ and $80\times1\times1$ elements. The microplane stress integration is performed by a Voronoi integration scheme (See Appendix A) with 66 microplane over the entire unit sphere. In order to control the location of the the localization zone, the central one tenth of the bar is weakened by reducing the tensile strength by 10 \%.

For comparison, simulations with $r_0=0$ are also performed. As one might expect, for $r_0=0$ the bar response is mesh dependent. The load displacement curve is increasingly more brittle (Fig.~\ref{reg_uniaxial_bar}a) as the mesh become more refined and damage localizes always in one single element as one can see in Fig.~\ref{reg_uniaxial_bar}e where the longitudinal strain profile is plotted for a bar end displacement of 0.2 mm. Figs.~\ref{reg_uniaxial_bar}b and Fig.~\ref{reg_uniaxial_bar}f show that mesh sensitivity is only partially mitigated for $r_0\neq0$. In this case, the bar response is increasingly brittle upon mesh refinement immediately after the peak. The solution is less dependent on the mesh in the softening branch. The strain profile (Fig.~\ref{reg_uniaxial_bar}f  does not show the extreme strain localization of the local simulation but still there is not a clear convergence upon mesh refinement.

Cusatis and Zhou \cite{cusatis2013high} conducted a 1D spectral analysis of localization and proposed to regularize the solution by computing the high order stresses as
\begin{equation}
\label{eq:Reqhigherorderstress}
\Sigma_{ijk}=\frac{3r_0}{4 \pi} \int_{\Gamma_1} \left(\sigma_{\beta}+ S_0 \psi_{\beta}\right) \mathcal{P\hspace{0.2mm}}_{ijk}^{\beta} \mathrm{d}S
\end{equation}
where $\psi_{\beta}=r_0 \mathcal{P\hspace{0.2mm}}_{ijk}^{\beta} \Gamma_{ijk}$ and $S_0$ is the localization limiter. For the 1D case and for linear softening, one can obtain a localization band equal to $2 r_0$ if $S_0=\left(1 + 1/ \pi^2 \right)H_t $ \cite{cusatis2013high}. For exponential softening, in which the softening modulus is not constant, the localization limiter must be a function of the current value of strain. In this case, and again with reference to a 1D setting, one can write $S_0=\left(1 + 1/ \pi^2 \right)H_t \exp(-H_t \langle\epsilon-\epsilon_t \rangle)$ with $\epsilon_t=\sigma_t/E_0$. In order to use this 1D result in the microplane formulation one can write
\begin{equation}
S_0(\epsilon, \omega)=\left(1 + 1/ \pi^2 \right)H_0(\omega) \exp(-H_0(\omega) \langle\epsilon-\epsilon_0(\omega) \rangle)
\label{eq:localization-parameter}
\end{equation}
where $\omega$ accounts for the presence of shear on the microplanes (see the adopted constitutive equations above).

From the physical point of view, this regularization strategy corresponds to adding, in parallel with the softening microplane stresses, additional non-softening microplane stresses which ensure that the phase velocity of a propagating uniaxial wave is always real. It is worth noting that such parallel coupling of softening and elastic components is also intrinsically captured by the microplane model by virtue of the kinematic constraint (Eq. \ref{eq:def_microplane_strain}) and the micro-macro stress relation (Eq. \ref{eq:stress_integration_sphere}). This is the reason why partial regularization is attained with $r_0 \neq 0$ and $H_0=0$ (Figs. \ref{reg_uniaxial_bar}b and \ref{reg_uniaxial_bar}f)

Figs.~\ref{reg_uniaxial_bar}c and ~\ref{reg_uniaxial_bar}g show the obtained response by adopting the formulation in Eqs. \ref{eq:Reqhigherorderstress} and \ref{eq:localization-parameter}. As one can see, the initial softening branch is very well regularized. However, the rest of the softening curve shows an oscillatory behavior, the convergence upon mesh refinement is not clear, and a residual load appear even for relatively large strains (stress locking).  The strain profile does not localize in one element but, again, clear convergence cannot be demonstrated. 

The stress locking phenomenon is due to the fact that the additional elastic stresses associated with $H_0$ are added during the entire simulation, since the beginning when the material has yet to soften. This is clearly unnecessary. An alternative regularization scheme which avoids this shortcoming can be formulated by introducing the regularization term on the stress increments as opposed to total stress. In this case Eq. \ref{eq:Reqhigherorderstress} is substituted by the following  equations

\begin{equation}
\label{eq:Reqhigherorderstress-rate}
\mathrm{d}  \Sigma_{ijk}=\frac{3r_0}{4 \pi} \int_{\Gamma_1} \left( \mathrm{d}  \sigma_{\beta}+S_0 \mathrm{d}  \psi_{\beta}\right) \mathcal{P\hspace{0.2mm}}_{ijk}^{\beta} ds; ~~~\Sigma_{ijk}=\int   \mathrm{d}  \Sigma_{ijk}
\end{equation}
where $S_0=0$ for $\varepsilon < \varepsilon_0(\omega)$ and Eq. \ref{eq:localization-parameter} holds for $\varepsilon \geq \varepsilon_0(\omega)$.

Figs. \ref{reg_uniaxial_bar}d and \ref{reg_uniaxial_bar}h shows the results obtained with this incremental formulation. As one can see, both the load displacement curve and the strain profile along the bar are fully regularized and both show a clear convergence upon mesh refinement. The localization band is larger than the theoretical 1D value of $2r_0$. This is due to the intrinsic 3D formulation of the microplane model which, even under uniaxial tensile macroscopic conditions, feature tensile and shear strains and stresses at each microplane orientation. The correct estimate of the localization band requires a fully 3D spectral analysis, which, however is intractable from the analytical point of view (see discussion in Ref. \cite{cusatis2013high}).


\section{Conclusions} 

This paper presents the isogeometric finite element implementation of a recently developed High-Order Microplane (HOM) model. The HOM model was originally derived based on a discrete particle model and the resulting theory  includes gradient elasticity and Cosserat theory as a special cases. In addition, the HOM model allows for an effective regularization of softening constitutive equations by means a simple modification of the high order stresses. The numerical simulations are carried out with an isogometric finite element characterized by 27 control points and 6 degrees of freedom, 3 displacement components and 3 rotation components, at each control point. The spatial integrals were performed with a gauss quadrature scheme with 27 gauss points and the microplane integration was carried out by a novel integration scheme based on the Voronoi discretization of the unit sphere with 66 microplanes.

Based on the results presented in the paper the following conclusions can be drawn: 

 \begin{enumerate}
 \item The implemented finite element with adoption of the HOM model as constitutive equation performs very well in the numerical simulations of classical high order elasticity problems. This was verified by simulating bending of strain-gradient elastic cantilever beams and computing the stress concentration due to a circular hole in a Cosserat elastic plate subject to tension.
 \item The HOM formulation with softening constitutive laws is only partially regularized and it shows mesh dependence in the softening branch close to the peak load.
 \item The regularization technique previously proposed bt Cusatis and Zhou \cite{cusatis2013high} solves the mesh dependency in the initial post peak but leads to stress locking and lack of clear convergence in the far post peak.
\item An optimal regularization of the post peak response is obtained by introducing the regularizing term on the high order stress increments as opposed to the total high order stresses. With this approach a clear convergence can be observed for both the load versus displacement curve and the strain profile along the bar. 
\end{enumerate}

\section{Acknowledgment}
Financial support from the U.S. National Science  Foundation  (NSF) under grant  CMMI-1435923 is gratefully acknowledged. The work of the first author was also partially supported by the Scientific and Technological Research Council of Turkey (TUBITAK).


%
% ...


\appendix       
\section*{Appendix A: Voronoi Integration Scheme}

\begin{figure}[h!]
    \centering
    \includegraphics[width=0.5\textwidth]{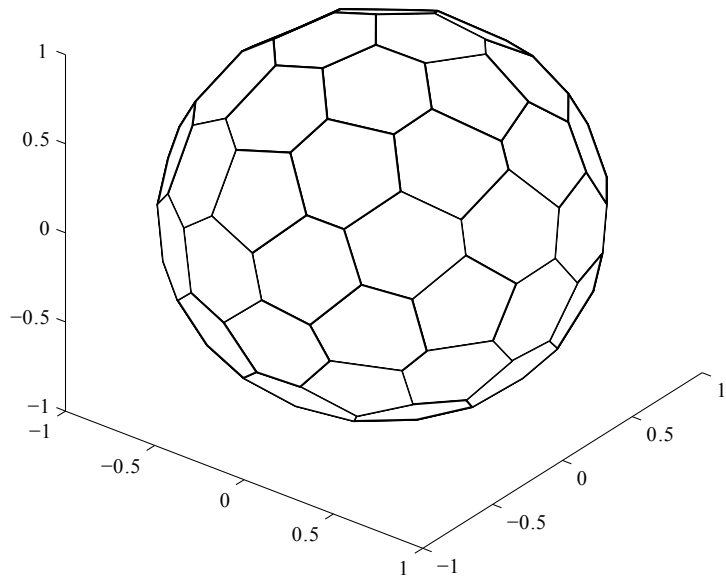}
    \caption{Voronoi tessellation of unit sphere with 66 facets}
    \label{fig:vorinoi66}
\end{figure}

For inelastic consitutive equations, Eq.  \ref{eq:stress_integration_sphere} can only be computed numerically. In previous microplane model work \cite{Bavzant1983Microplane,Bavzant1985Microplane,caner2013microplane} gaussian optimal integration formulas were developed for the integration over the unit hemisphere. However, for the high order microplane model adopted in this paper, integration over the entire unit sphere is required. Hence, a new integration scheme was developed based on the voronoi tessellation of the unit sphere \cite{}. Fig.~\ref{fig:vorinoi66} show the discretization of a unit sphere with 66 microplanes. The corresponding areas and spherical angles are reported in Table~\ref{table:microplanes66}

The local system of refence, attached to each microplane, can be calculated as $n_1=\sin\phi\cos\theta$, $n_2= \sin\phi\sin\theta$, $n_3=\cos\phi$;  $m_1=\cos\phi\cos\theta$,  $ m_2= \cos\phi\sin\theta$, $m_3=-\sin\phi$ ; and $l_1=-\sin\theta$,  $ l_2=\cos\theta$, $l_3=0$.

For the discretized sphere the energetical equivalence, that relates microplane stresses, microplane couple stresses, and microplane high order stresses to their macroscopic counterparts, can be written as
\begin{equation}
\sum_{s=1}^{n_p} ( \sigma_{ij} \delta \gamma_{ij} + \mu_{ij} \delta \kappa_{ij} + \Sigma_{ijk} \delta \Gamma_{ijk}) V_s = \sum_{s=1}^{n_p} R ( \sigma_\beta^s \delta \varepsilon_\beta^s + \mu_\beta^s \delta \chi_\beta^s) A_s
\label{Eq:PVW-discrete}
\end{equation}
where $n_p$ is the number of microplanes, $R=1$, $A_s$ is the area of the generic microplane and $V_s$ is the volume of a pyramid of unit eight and base area equal to $A_s$. Since the macroscopic quantities are uniform inside the unit sphere and by using the kinematic contraint in Eq. \ref{eq:def_microplane_strain}, one obtains
\begin{equation}
\sigma_{ij} = 3\sum_{s=1}^{n_p} w_s\sigma_\beta^s P_{ij}^s; ~~~\Sigma_{ijk} = 3\ell_0\sum_{s=1}^{n_p} w_s \sigma_\beta^s \mathcal{P}_{ijk}^s 
\label{Eq:PVW-stresses-discrete}
\end{equation}
where  $w_s=A_s/ \sum_{s=1}^{n_p} A_s$ is the area ratio of fmicroplane $s$ over the total area of all microplanes.

\clearpage

\begin{table}[t]
	\caption{Material parameters for the stress concentration simulations.}
	\begin{center}
		\label{table_SCF_mat}
		\begin{tabular}{c c c c c c c c}
			& & \\ 
			\hline
			Case & $N_{c}$ 	& $E$ [N/m$^2$] 			& $\nu \left[ - \right]$ & $\chi$ [N/m$^2$]& $\pi_1$ [N] & $\pi_2$  [N] & $\pi_3$ [N]\\
			\hline
			SIM1 & 0.00 	& 1.195E8	& 0.300	& 0.0		& 0		& 7.593E8		& 0	\\
			SIM2 & 0.25		& 1.263E8	& 0.308	& 6.895E6	& 0		& 8.541E9		& 0	\\
			SIM3 & 0.50 	& 1.856E7	& 0.346	& 6.895E6	& 0		& 1.708E9		& 0	\\
			SIM4 & 0.70		& 1.856E7	& 0.346	& 3.448E8	& 0		& 1.708E9		& 0	\\
			\hline
		\end{tabular}
	\end{center}
\end{table}

\begin{table}[t]
\caption{Calculated and analytical stress concentration factors.}
\begin{center}
\label{table_SCF}
\begin{tabular}{l c c c c c  }
& & \\ 
\hline
Case &  Analytical 	& Ref.~\cite{pothier1994three}		& Ref.~\cite{sachio1984finite}		& Mesh-1 & Mesh-2 \\
\hline
SIM1 &  3.0			& 3.0		& 2.99		& 2.894		& 2.999	\\
SIM2 &  2.839		& 2.86		& 2.83		& 2.735		& 2.813		\\
SIM3 &  2.482		& 2.58		& 2.67		& 2.436		& 2.474		\\
SIM4 &  2.034		&  -			&  -			& 2.011		& 2.034		\\
\hline
\end{tabular}
\end{center}
\end{table}

\begin{table}[t]
\small
	\caption{Spherical angels and weight for the Voronoi integration scheme on the unit sphere with 66 microplanes.}
	\begin{center}
		\label{table:microplanes66}
		\begin{tabular}{c c c c c c c c}
			& & \\ 
			\hline
			Index & $\phi$ [rad]	& $\theta$ [rad] & $w$ & Index & $\phi$ [rad] & $\theta$ [rad]  & $w$\\
			\hline
1 &   1.33030045275 &   5.00374576373 &   0.01556289035 & 34 &   2.31088950846 &   4.82219824701 &  0.015659816774 \\
2 &   1.80674103963 &   4.85180287262 &  0.015525072181 & 35 &   1.68356115574 &   3.62959310171 &  0.012703614433 \\
3 &   1.83428166859 &   2.37761582873 &  0.015772995694 & 36 &   1.17908789017 &   2.86363192111 &  0.015494119892 \\
4 &  0.278016102851 &    3.5135300969 &  0.016411951508 & 37 &   2.43326894062 &   4.17259028347 &   0.01585851993 \\
5 &   1.19867810799 &   6.05790630937 &  0.015959860033 & 38 &  0.751991114512 &   3.74125121886 &  0.015542679034 \\
6 &   2.18666735888 &    6.1851151041 &  0.015766111972 & 39 &   1.48174700087 &   4.52738749074 &  0.016234007126 \\
7 &   2.51505669679 &   1.00948941577 &  0.015815722278 & 40 &   1.17797319071 &   5.47732574618 &  0.015237652047 \\
8 &   1.97044646628 &   4.39097031362 &   0.01620127622 & 41 &   2.75538217785 &    5.0325964137 &  0.015882225302 \\
9 &   2.22482662721 &  0.504094970079 &  0.015706318065 & 42 &   1.60511144111 &   4.07892986465 &  0.016271016825 \\
10 &  0.793517629239 &   5.79647595823 &  0.015848007878 & 43 &  0.211433971236 &   1.04690369381 &   0.01534134374 \\
11 &   1.39809888725 &   1.58553021733 &  0.015599630829 & 44 &   1.20096747392 &   0.69784980578 &   0.01603447883 \\
12 &  0.739067275481 &  0.769739813641 &  0.015440641966 & 45 &   1.86276085468 &  0.260411929155 &  0.012502928389 \\
13 &  0.961985366598 &   0.21029809686 &  0.016174983693 & 46 &   2.37643960157 &   3.40278966551 &  0.016041396391 \\
14 &  0.950437303827 &   2.37050395082 &  0.015942151322 & 47 &   1.62207830068 &   5.32084284346 &  0.015259454437 \\
15 &  0.813625026449 &   5.13504832577 &   0.01543535868 & 48 &   2.61760888454 &   1.83810114382 &   0.01257814012 \\
16 &   2.61145274129 &   2.69145782712 &  0.015194499446 & 49 &   2.16892121481 &   2.77080563998 &  0.015801726157 \\
17 &   2.23497522046 &   2.15245254337 &  0.015345512032 & 50 &   2.65416114082 & 0.0586808280569 &  0.015886144057 \\
18 &  0.514064131767 &   2.20184861497 &  0.015932392513 & 51 &   1.52857750031 &   5.74473287538 &   0.01250742307 \\
19 &   1.94829281362 &   3.26134434253 &  0.015825043399 & 52 &   3.01082211471 &   1.19566768512 &   0.01518447752 \\
20 &    1.4105201288 &  0.249853752915 &  0.015616800339 & 53 &   1.39454341264 &   2.05400049633 &  0.015767805536 \\
21 &   2.05496644773 &   5.24902003738 &  0.012703348341 & 54 &   1.96388773167 &   5.72848024719 &  0.015148298878 \\
22 &  0.520612112758 & 0.0922663976576 &  0.012851408141 & 55 &  0.691607842915 &   2.95351448068 &  0.015670054589 \\
23 &   1.25290443237 &   3.73412539672 &  0.015226387285 & 56 &   1.41041577607 &   2.47844583949 &  0.013231436974 \\
24 &   2.80827281345 &   3.67831502457 &   0.01274928792 & 57 &    1.6890059202 &   6.14390402236 &  0.015532115173 \\
25 &   1.79525012998 &   1.40629465268 &  0.012691306532 & 58 &   1.81739120815 &   1.87940062947 &  0.015805899519 \\
26 &   2.06509258359 &   3.86211165764 &   0.01559038684 & 59 &  0.623448397599 &   4.42059901363 &  0.015983191096 \\
27 &   1.69358920808 &   2.84141732452 &   0.01567310381 & 60 &   2.42095409404 &   5.59648563043 &  0.015886463625 \\
28 &   1.10513075433 &   4.17848693448 &  0.015600274168 & 61 &   1.68856577065 &  0.663758643732 &  0.015895401615 \\
29 &  0.358830558297 &    5.2784396103 &  0.015929283478 & 62 &  0.652211552292 &   1.43220215503 &  0.012532311844 \\
30 &   2.02914773443 &   1.00176517068 &  0.015817892577 & 63 &   2.21418943469 &   1.54680613956 &  0.015534310432 \\
31 &   1.05116632586 &   4.65626133789 &  0.012697330999 & 64 &   1.51910545537 &   1.07794424578 &  0.015511054514 \\
32 &  0.984415156942 &   1.80817356035 &  0.015699358498 & 65 &   1.04178071878 &   3.32048914291 &  0.012279112331 \\
33 &   1.47638971563 &   3.24915411931 &  0.015478315139 & 66 &   1.07810787109 &   1.22571599163 &  0.015416475676 \\
			\hline
		\end{tabular}
	\end{center}
\end{table}

\clearpage

\begin{figure} 
    \centering	
	\includegraphics[width=0.25\linewidth] {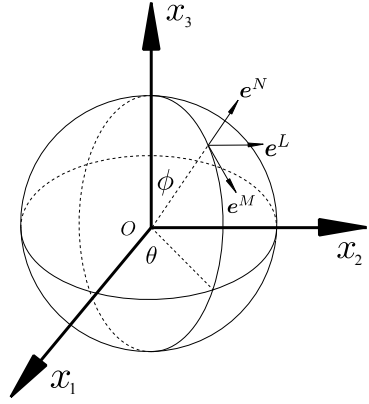}
		\caption{Global and microplane system of references.}
	\label{geometry}
\end{figure}  

\begin{figure} 
    \centering
	\includegraphics[width=0.4\linewidth] {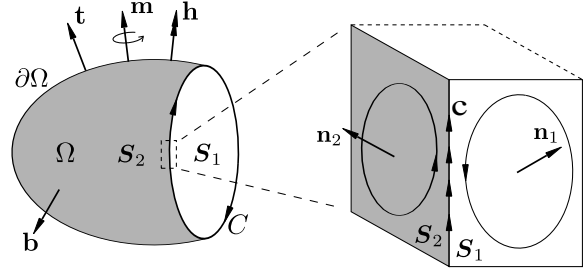}
	\caption{Boundary conditions for the high order microplane model.}
	\label{HOM_BC}
\end{figure} 

 \begin{figure}
 	\centering	
 	\begin{subfigure}{.33\textwidth} 		
 		\includegraphics[width=1\linewidth] {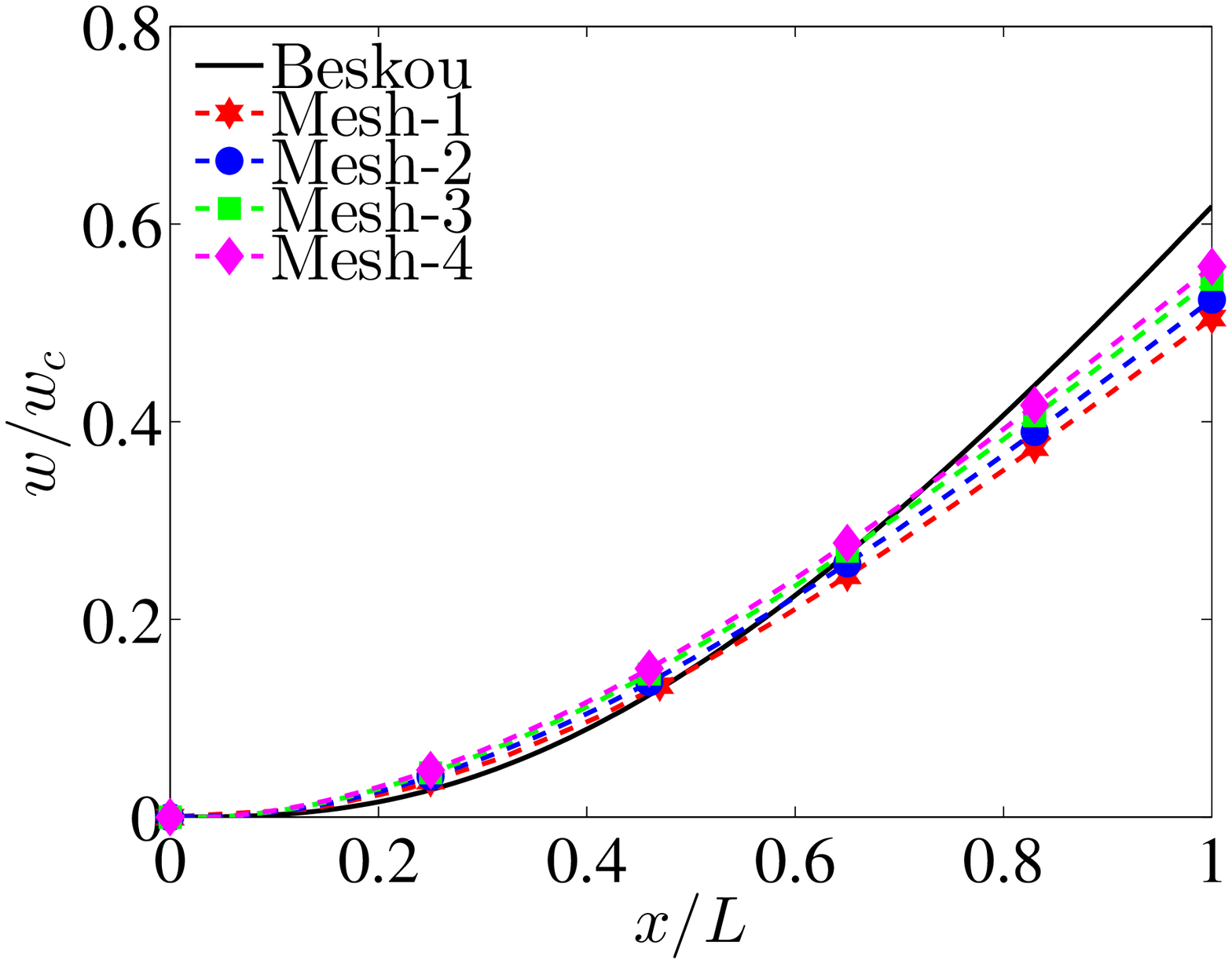}
 		\caption{}
 		\label{fig:cantilever_beam_con}
 	\end{subfigure}%
 	\begin{subfigure}{.33\textwidth} 		
 		\includegraphics[width=1\linewidth] {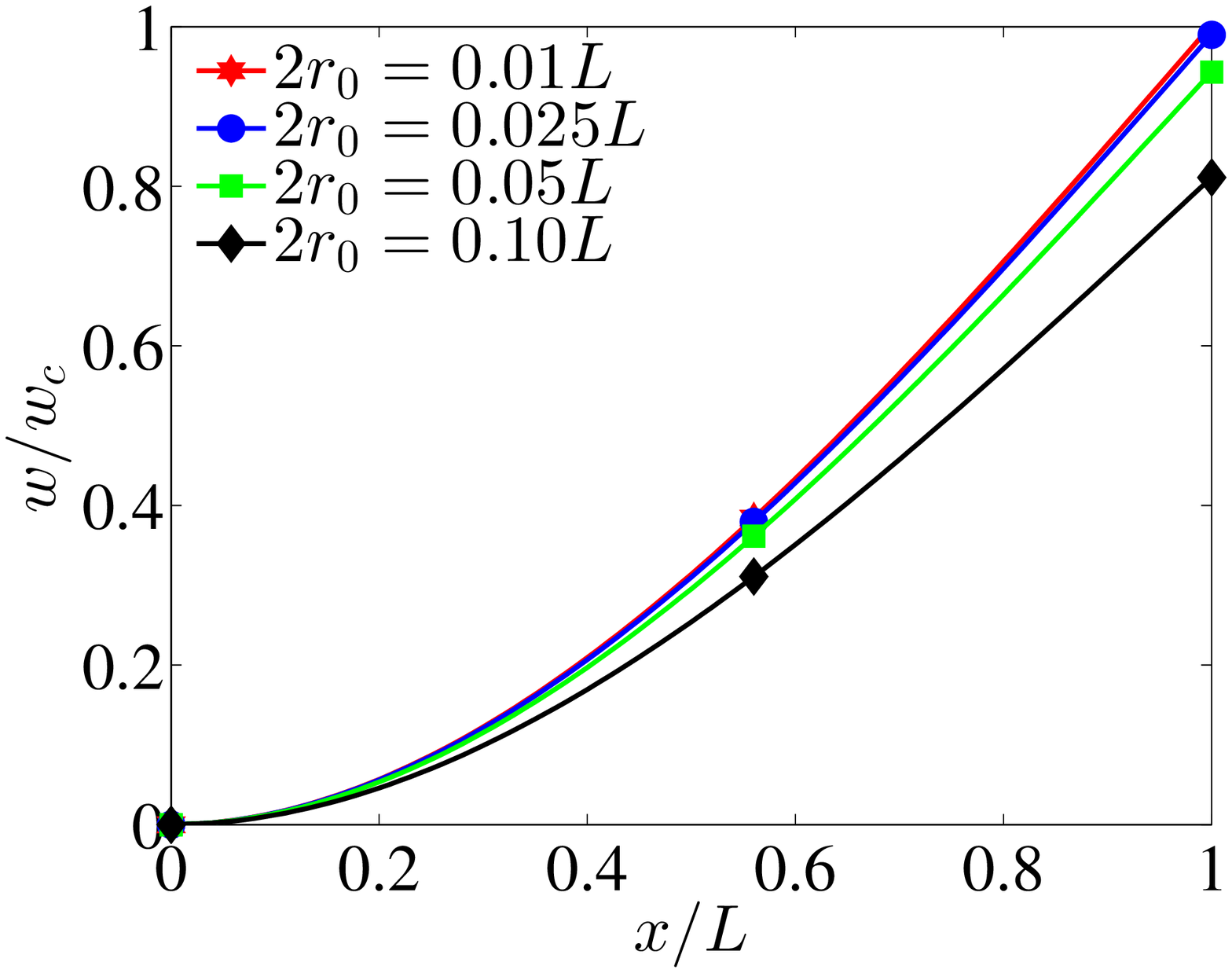}
 		\caption{}		
 		\label{fig:cantilever_beam_disp}
 	\end{subfigure} 
 	\begin{subfigure}{.33\textwidth} 		
 		\includegraphics[width=1\linewidth] {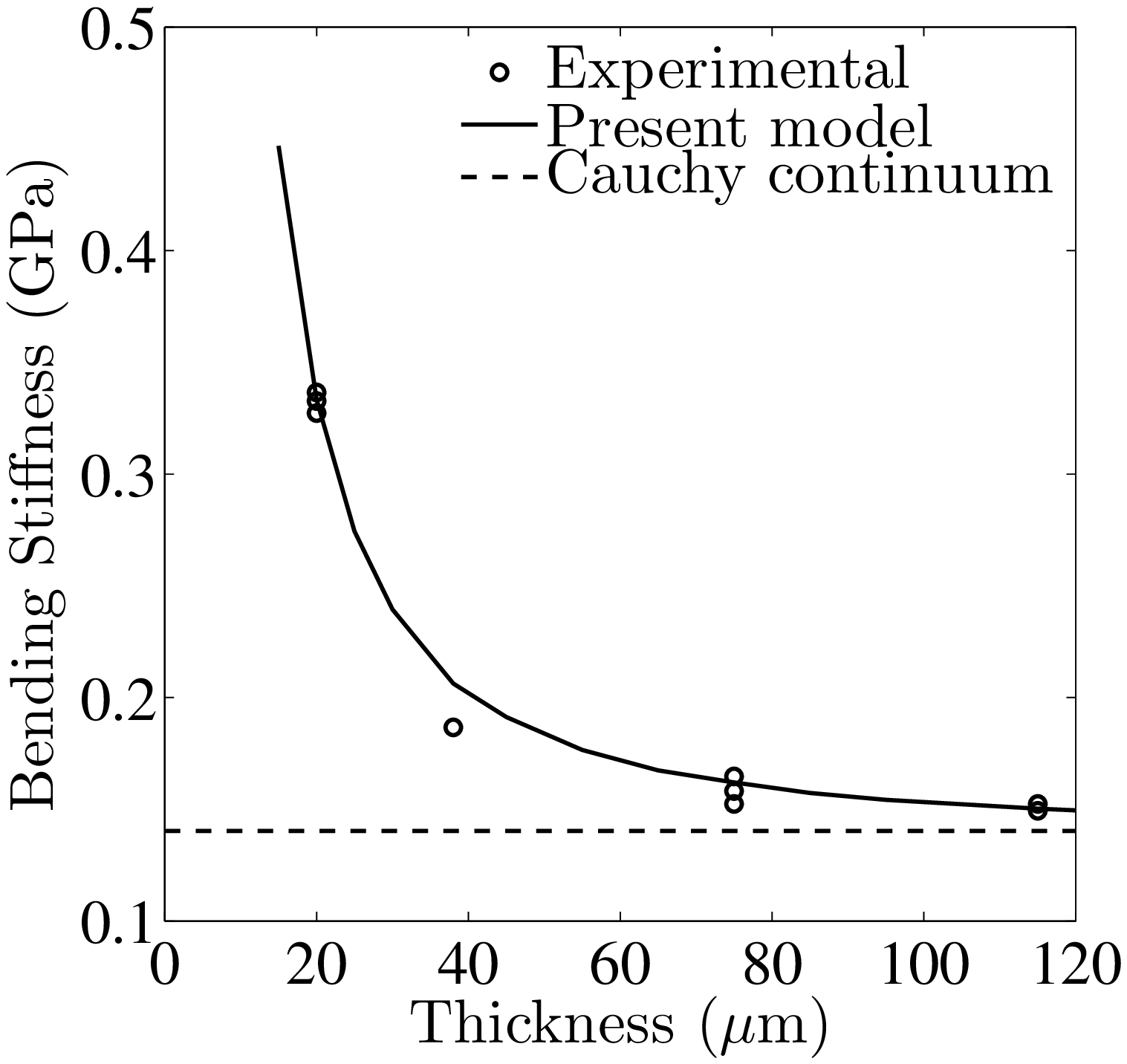}
 			\caption{}
 			\label{fig:cantilever_beam_stiffness}
 		\end{subfigure}%
 	\caption{Numerical results for the cantilever beam problems. (a) Elastic curves for various meshes and comparison with beam theory solution. (b) Calculated response for different internal length values.}	
 	\label{cantilever_beam}
 	\end{figure}

  \begin{figure}
 	\centering	
 	\begin{subfigure}{.33\textwidth} 		
 		\includegraphics[width=1\linewidth] {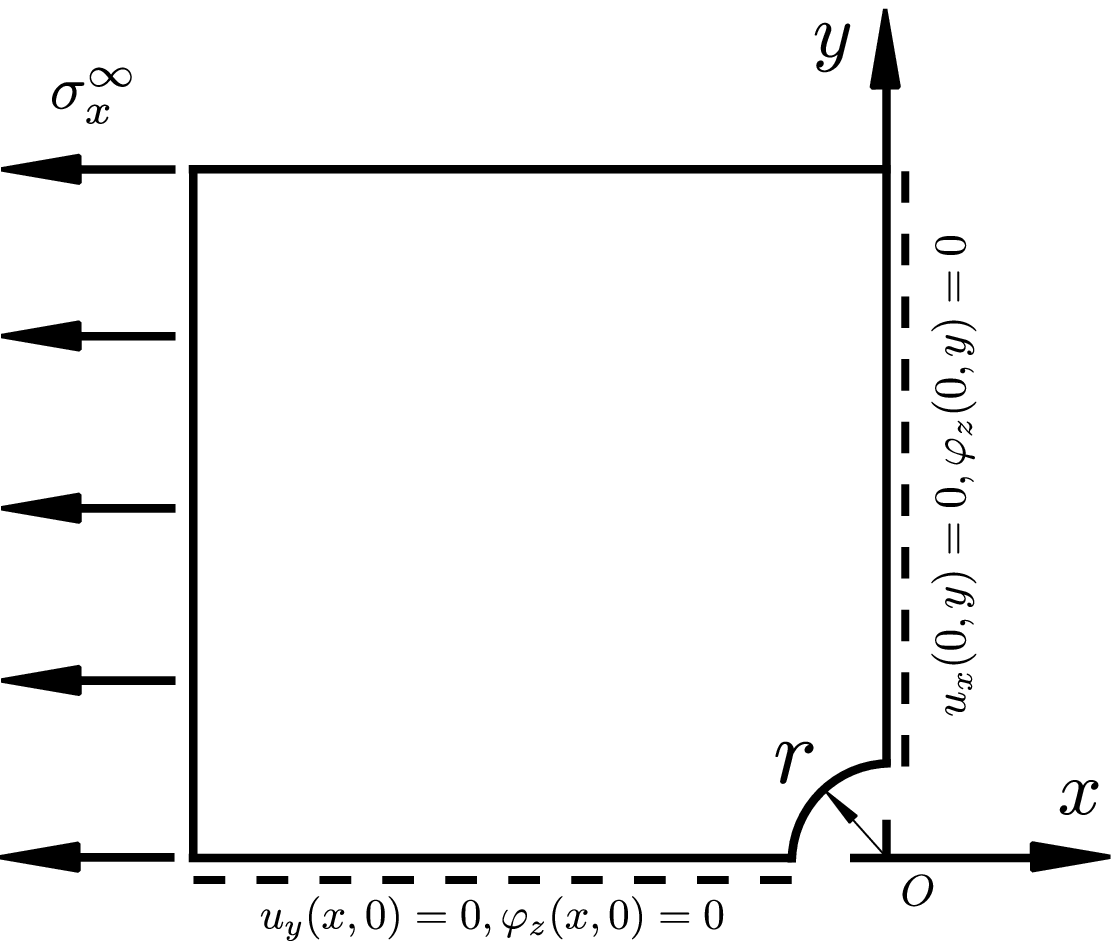}
 		\caption{}
 		\label{fig:iga_plate}
 	\end{subfigure}%
 	\begin{subfigure}{.5\textwidth} 		
 		\includegraphics[width=1\linewidth] {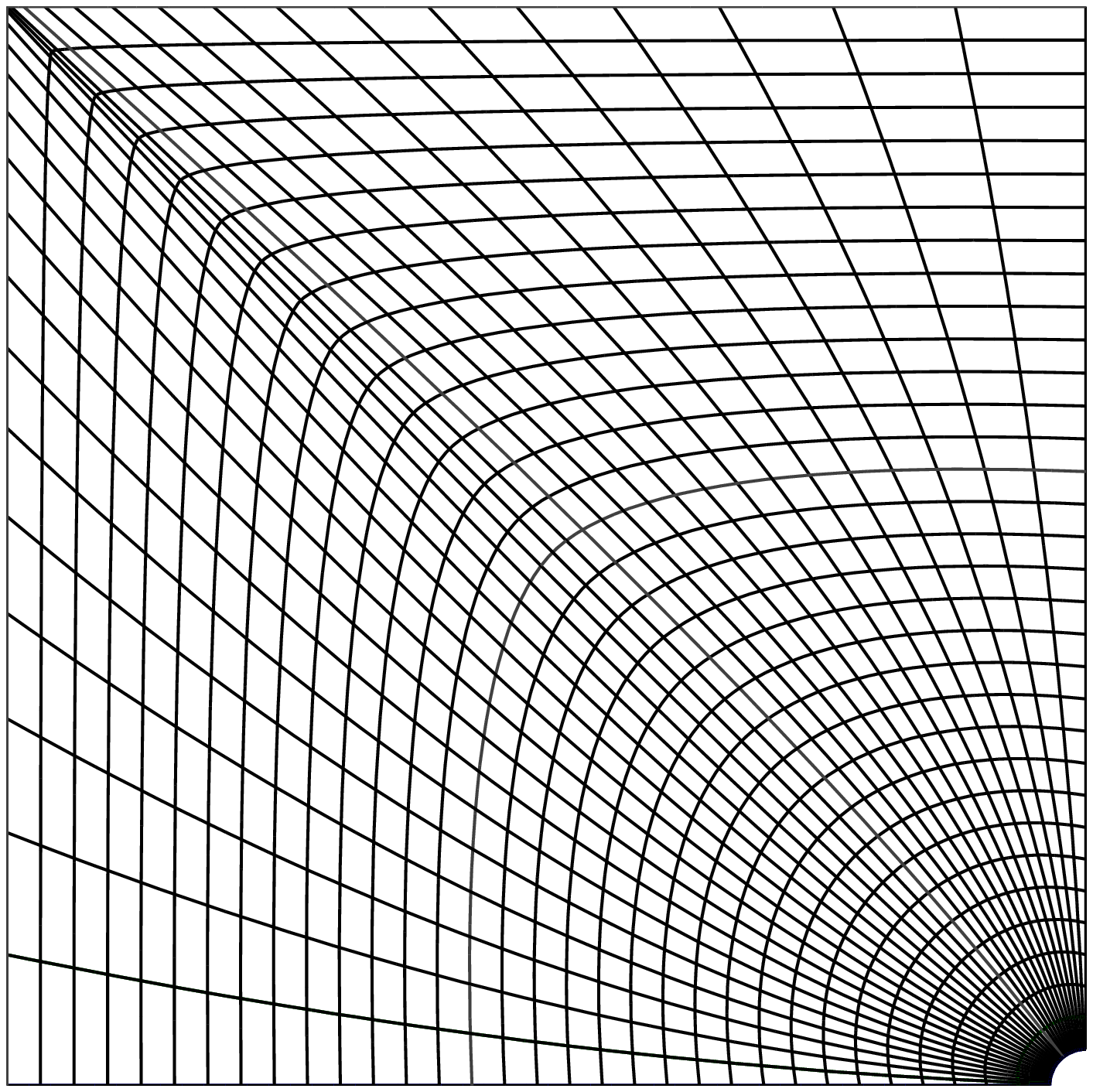}
 		\caption{}		
 		\label{fig:iga_mesh}
 	\end{subfigure} 
 	\begin{subfigure}{.15\textwidth}	 		
 		\includegraphics[width=1\linewidth] {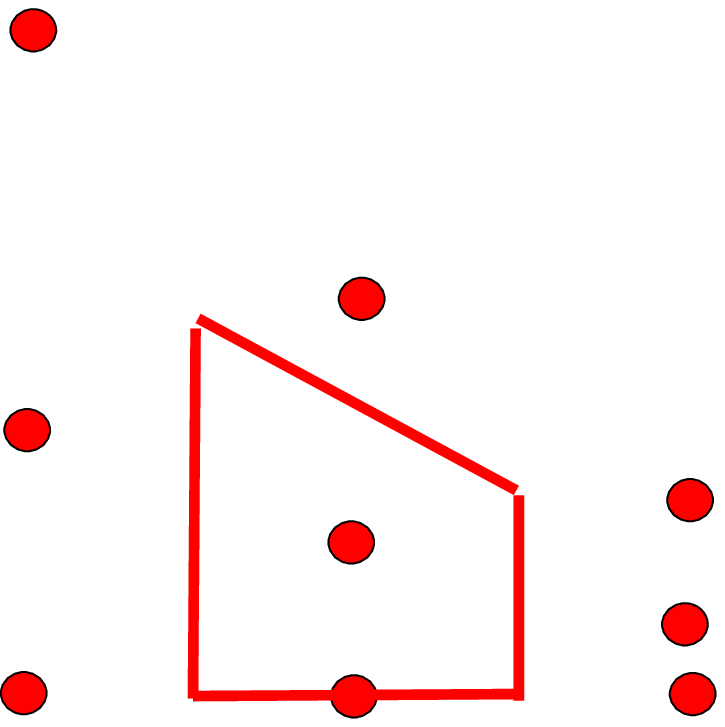}
 			\caption{}
 			\label{fig:iga_mesh_zoom}
 		\end{subfigure}%
\caption{Plate with a hole under tension. (a) Schematic sketch of loading and boundary conditions. (b) Typical finite element mesh. (c) Mesh zoom-in with highlited isogeometric elelemnt with corresponding control points.}
\label{plate_hole} 	
\end{figure}

\begin{figure}
	\centering	
	\begin{subfigure}{.24\textwidth}		
		\includegraphics[width=1\linewidth] {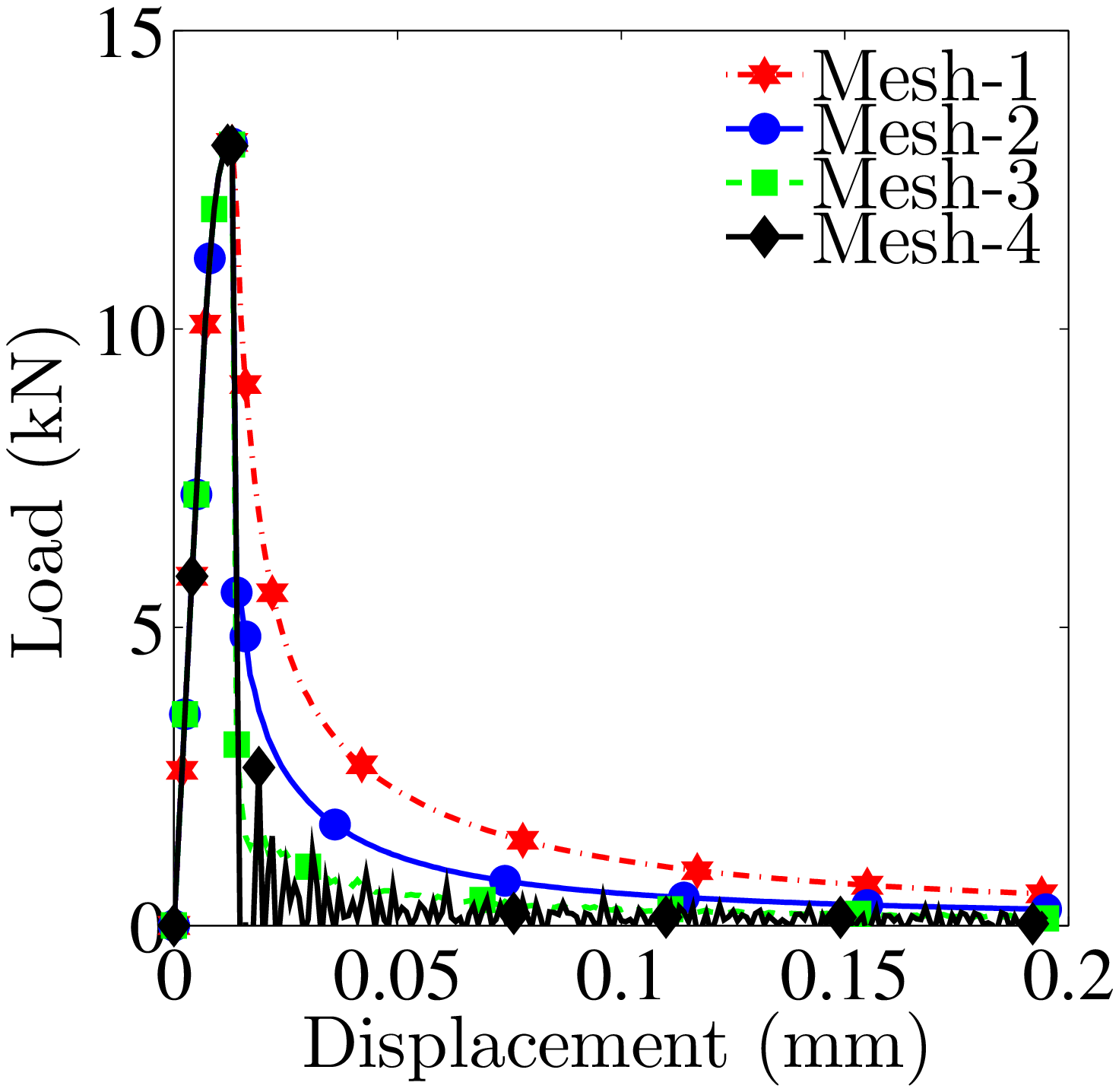}
		\caption{}
		\label{fig:LD_NOHOS}
	\end{subfigure}%
	\begin{subfigure}{.24\textwidth}		
		\includegraphics[width=1\linewidth] {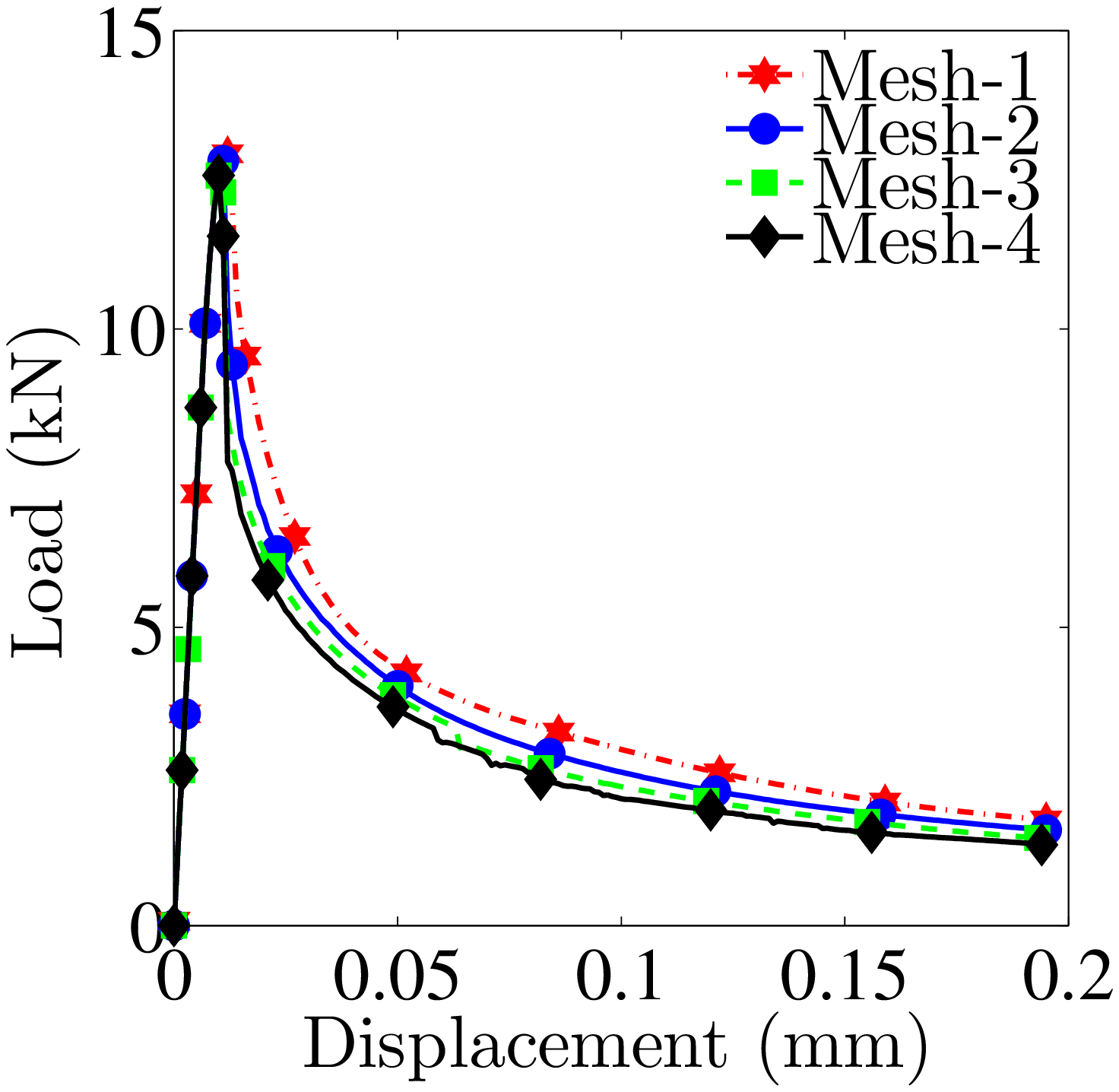}
		\caption{}		
		\label{fig:LD_NOREG}
	\end{subfigure} 
		\begin{subfigure}{.24\textwidth}			
			\includegraphics[width=1\linewidth] {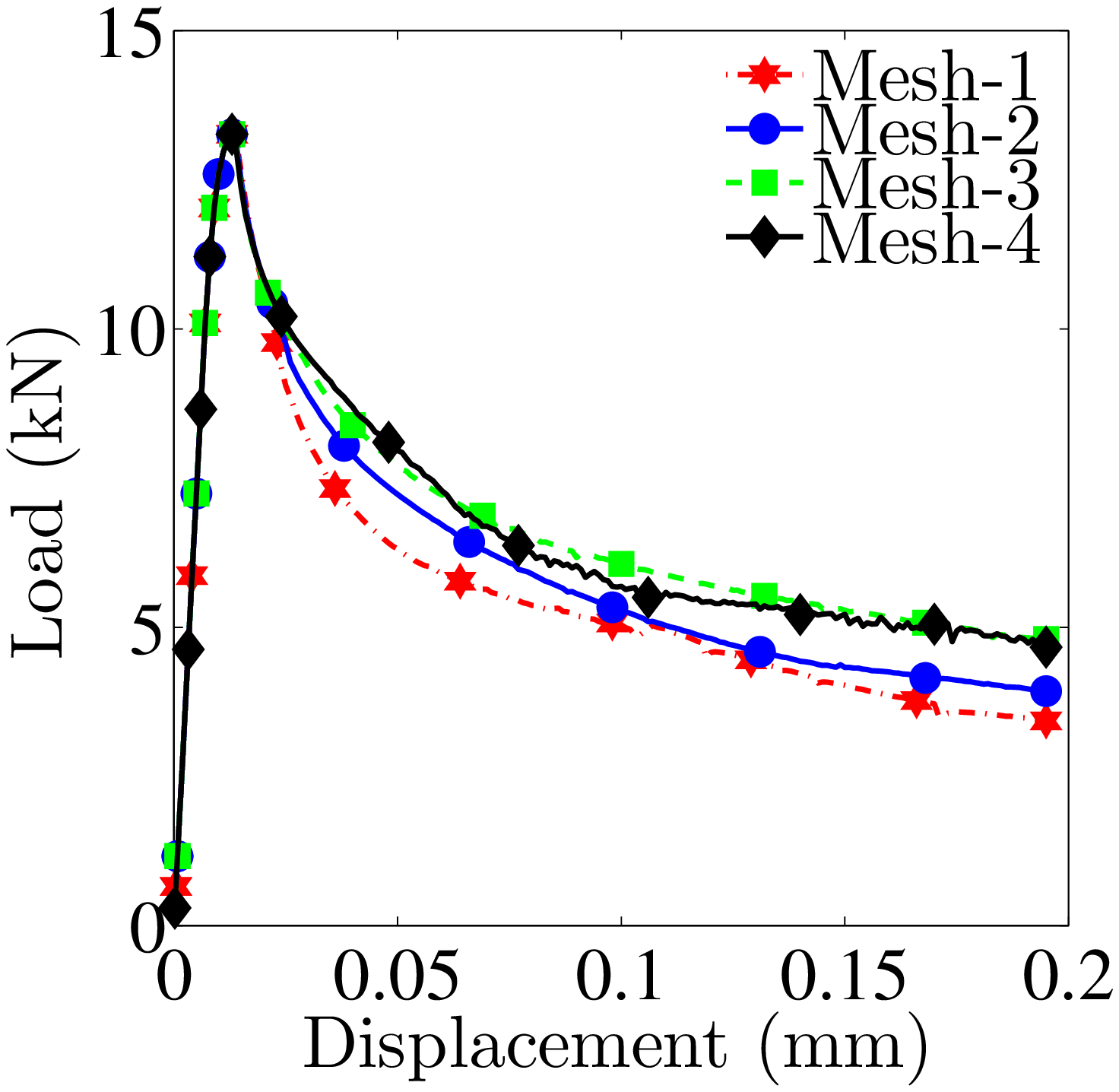}
			\caption{}
			\label{fig:LD_TOTALREG}
		\end{subfigure}%
		\begin{subfigure}{.24\textwidth}			
			\includegraphics[width=1\linewidth] {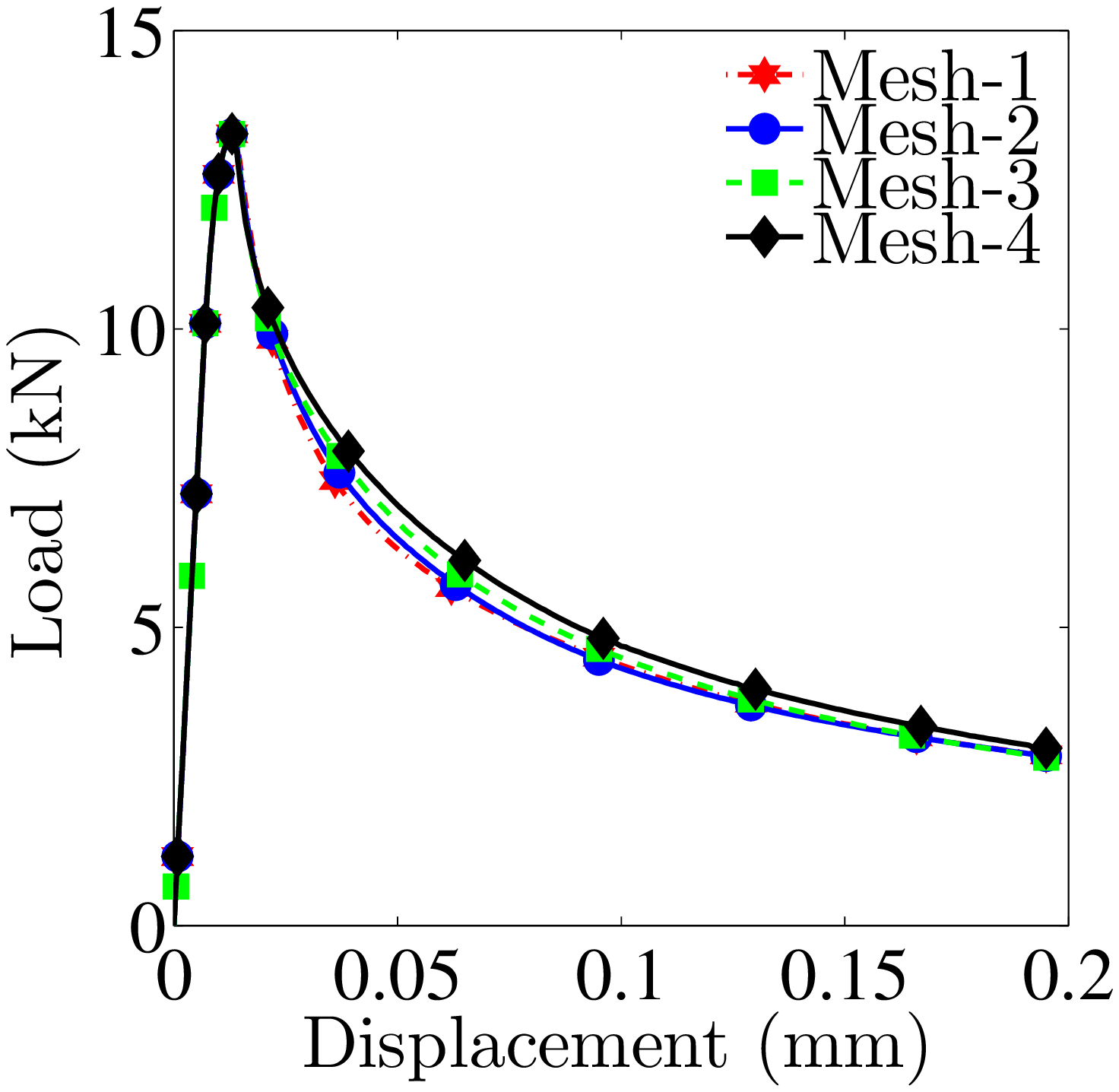}
			\caption{}		
			\label{fig:LD_REG}
		\end{subfigure} \\
			\begin{subfigure}{.24\textwidth}			
				\includegraphics[width=1\linewidth] {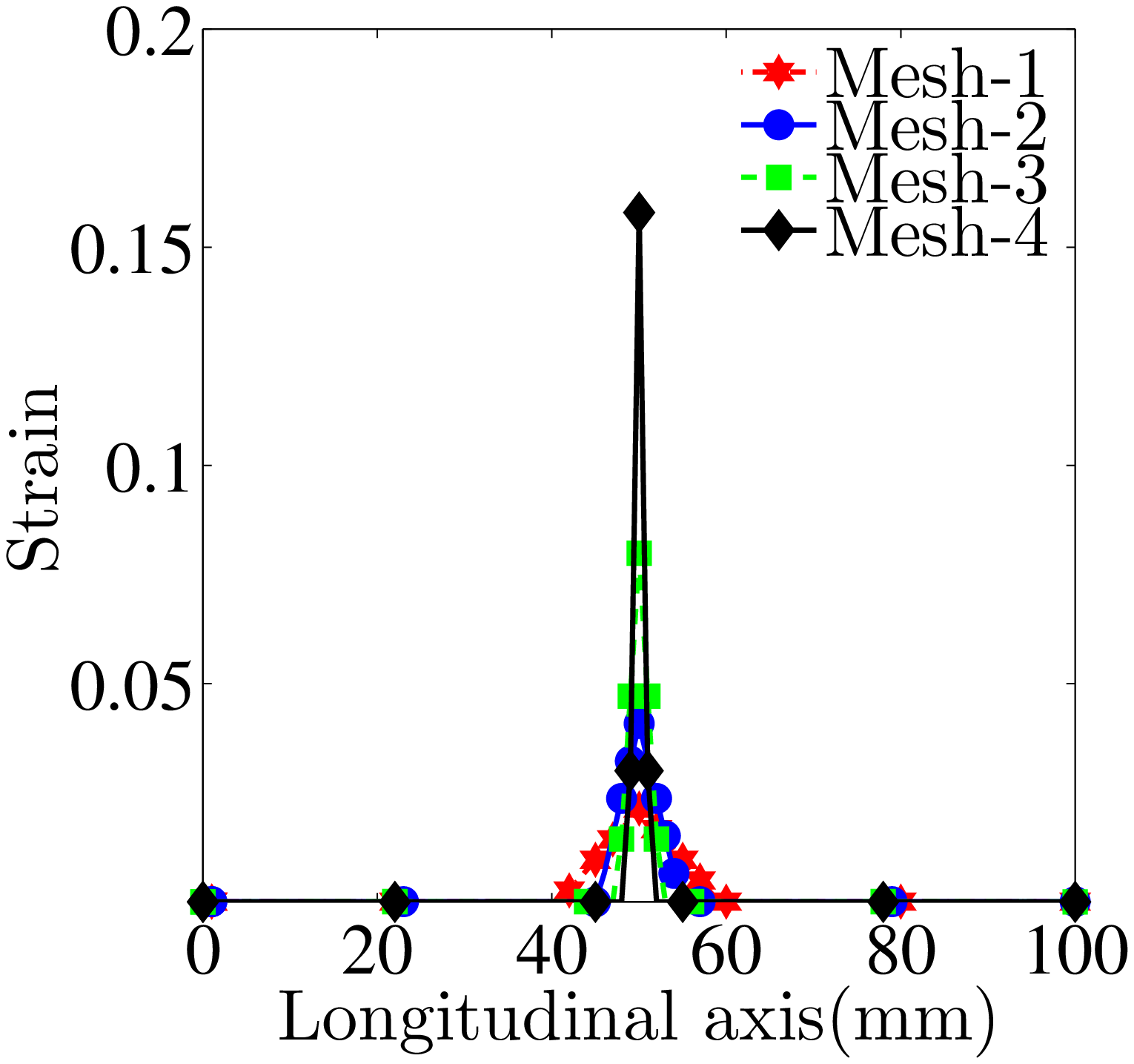}
				\caption{}
				\label{fig:Strain_NOHOS}
			\end{subfigure}%
			\begin{subfigure}{.24\textwidth}			
				\includegraphics[width=1\linewidth] {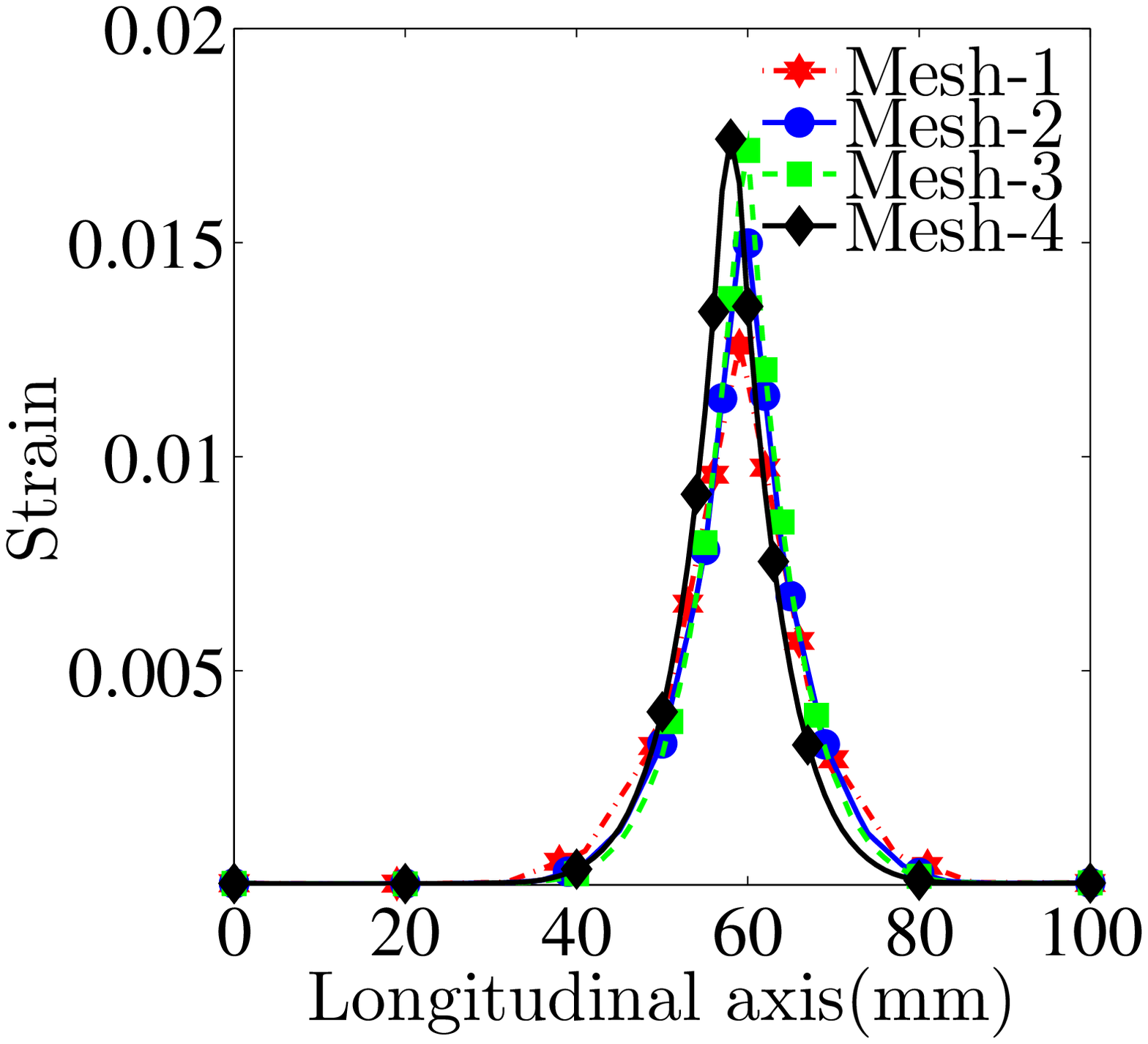}
				\caption{}		
				\label{fig:Strain_NOREG}
			\end{subfigure} 
			\begin{subfigure}{.24\textwidth}			
				\includegraphics[width=1\linewidth] {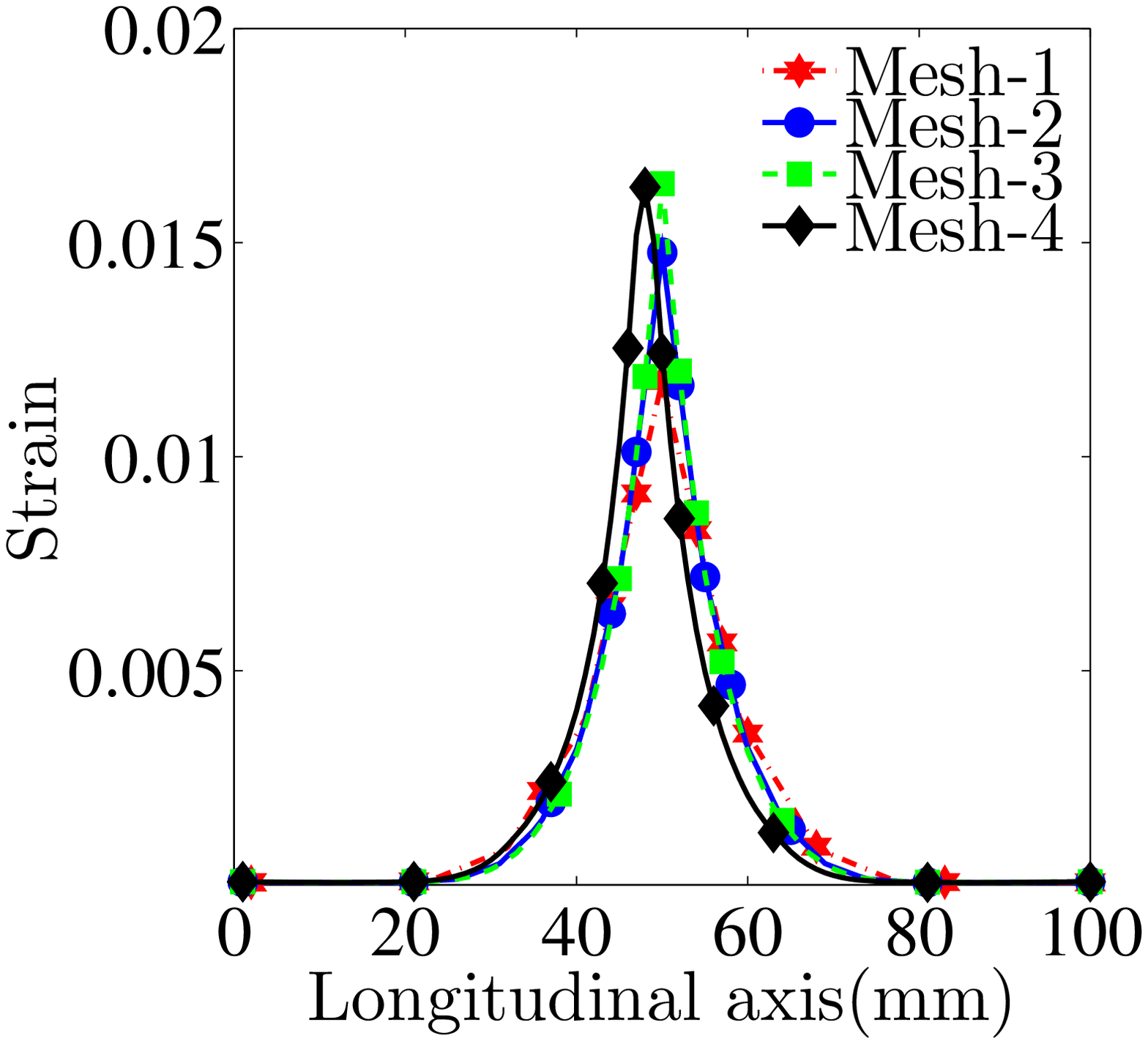}
				\caption{}
				\label{fig:Strain_TOTALREG}
			\end{subfigure}%
			\begin{subfigure}{.24\textwidth}			
				\includegraphics[width=1\linewidth] {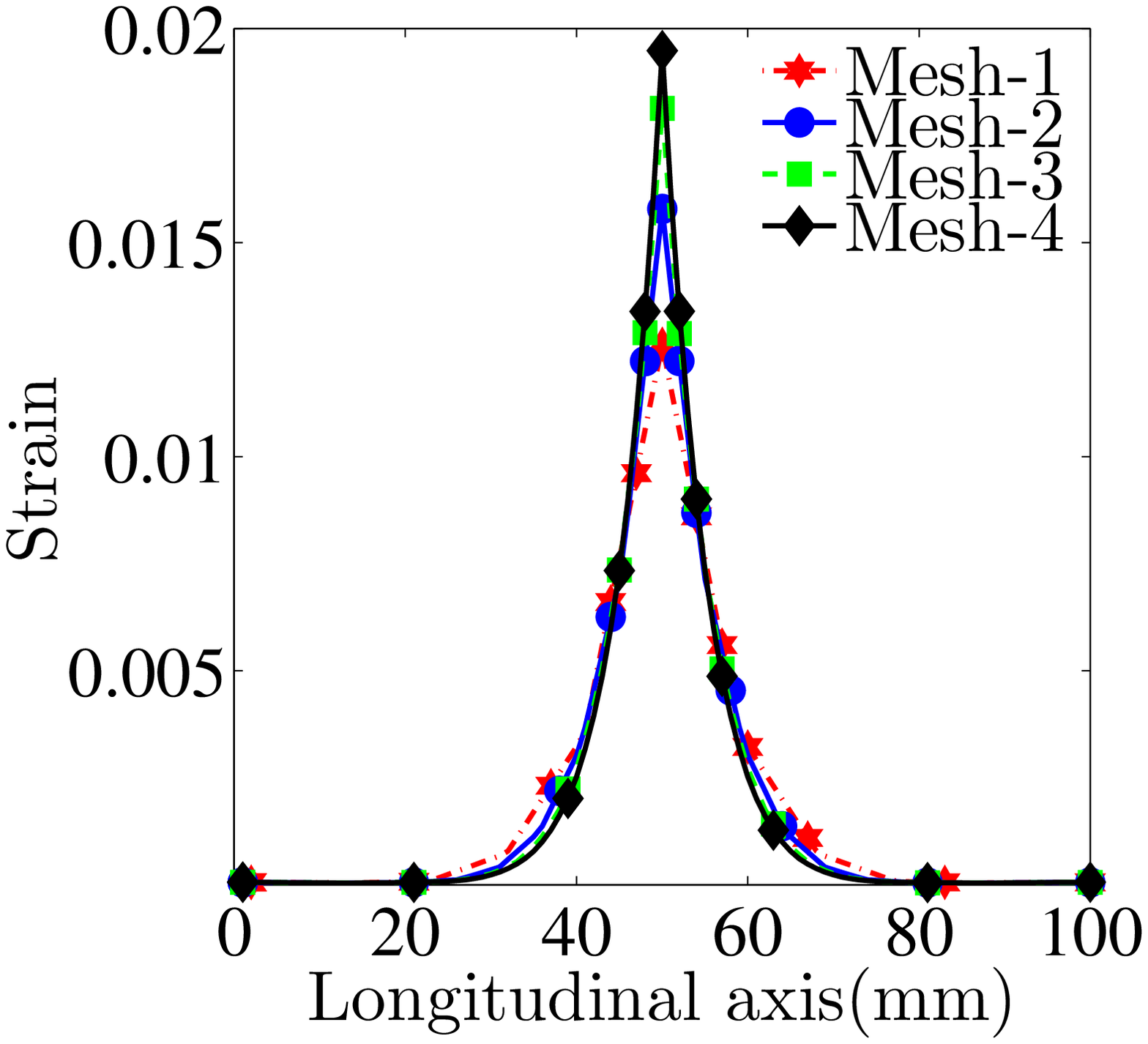}
				\caption{}		
				\label{fig:Strain_REG}
			\end{subfigure} 
			\caption{Load-displacement and strain profile curves for a bar under uniaxial tension: a) local formulation, b) high order formulation with no regularizion , c) high order formulation with regularization on the total high order stresses, d) high order formulation with regularization on the high order stresse increments.}	
			\label{reg_uniaxial_bar}
\end{figure}

\end{document}